\documentclass[aps,prl,onecolumn,groupedaddress,floatfix, 10pt]{revtex4-2}

\usepackage{float} 
\usepackage{hyperref} 
\hypersetup{colorlinks=true, urlcolor=blue} 
\usepackage{graphicx} 
\usepackage{mathtools} 
\usepackage{amssymb,bm} 
\usepackage{physics} 
\usepackage{xcolor} 
\usepackage{subfiles} 
\usepackage{lineno} 
\usepackage[sectionbib]{bibunits}
\newcommand{\degree}{$^\circ$} 
\usepackage{comment}

\newcommand{\angstrom}{\mbox{\normalfont\AA}} 
\newcommand{\pos}{\va{R}_{i}} 
\newcommand{\posprime}{\va{R}_{i'j'}} 
\newcommand{\Qion}{Q_{i}} 
\newcommand{\Qionprime}{Q_{i'}} 
\newcommand{\pion}{\va{p}_{i}} 
\newcommand{\pionprime}{\va{p}_{i'}} 
\newcommand{\Rh}{R_{h_2 - h_1}} 
\newcommand{\sumions}{\sum_{(i',j',\kappa') \neq (i,j,\kappa)}} 

\begin{document}

\begin{bibunit}

\title{Atomic and bond polarization causing strong screening of short-range Coulomb interactions and its effect in cuprate superconductors}
\author{Nassim Derriche}
\author{George Sawatzky}
\affiliation{Department of Physics and Astronomy \& Stewart Blusson Quantum Matter Institute,
	The University of British Columbia, Vancouver BC, Canada V6T 1Z4}

\begin{abstract}
	We present a novel and efficient real space, semiclassical model of electric polarization with general applicability to any system in which screening plays an important role. This model includes the effects of both atomic and bond polarizabilities, the latter originating from the modification of local bond charge transfer energies induced by polarizing charges. The nonlinear interference of multiple polarization clouds and the emergence of local field effects are highlighted as key phenomena highly influencing the short-range screening of the Coulomb interaction. As a representative system to showcase this model, the screened interaction between doped holes in the CuO$_2$ planes of cuprate high-temperature superconductors is investigated. This leads to the emergence of striking direction-dependent short-range minima in their Coulomb repulsion, which can strongly reduce the need for retardation effects and allow for an enhancement of the attractive interaction resulting from the exchange of bosons between two electrons or holes. This in turn enhances T$_C$, shortens the Cooper pair coherence length and supports the materialization of the pseudogap phase anisotropy observed in many high-T$_C$ superconductors.
\end{abstract}

\maketitle


\section{Introduction} 

In solids such as the transition metal compounds, it is well known that considerable electronic correlation effects are caused by strong short-range Coulomb interactions like the on-site repulsion of the $3d$ or $4f$ electrons which suppresses charge fluctuations, leading to localization and often resulting in magnetic insulators with ordered local magnetic moments \cite{cuprate_XPS_Mott, 4f_Mott_arpes}. It is also well known that these atomic Coulomb interactions, which have a bare energy scale of around 20 to 30 eV \cite{Cu_second_ionization_energy, Cu_ionization_potential}, are strongly reduced in a solid involving nearby strongly polarizable atoms \cite{anion_screening_sawatzky}. For example, on-site interactions in $3d$ transition metal compounds are effectively filtered down to between 5 and 10 eV due to the presence of highly polarizable O$^{2-}$ ions \cite{srvo3_hubbard_u, cuprate_sawatzky_xps_udd}. This very strong suppression raises the question of what the spatial landscape of the interactions between electrons on neighboring atoms in solids at a length scale of less than 20 \angstrom{} really is. In insulating solids such as many transition metal and rare earth compounds of great interest due to their wide range of tunable physical properties, a standard approach is to employ the Clausius-Mossotti method to relate the ionic polarizabilities of the constituent atoms to obtain an optical dielectric constant $\epsilon$ and then declare the screened interaction between electrons at positions $\va{r}_1$ and $\va{r}_2$ to be proportional to $\frac{1}{\epsilon |\va{r}_1-\va{r}_2|}$ \cite{Clausius_original, mossotti_original}. This corresponds to a monotonic decrease with distance. However, it has previously been shown in model calculations of short-range interactions that the polarizability of the constituent ions alone can lead to pronounced local minima in the energy landscape and to a strong dependence on the orientation of the two charges in the crystal lattice \cite{FeAs_Sawatzky, jeroen_screening_organic, Caprara_cuprate_cdw_anisotropy}. This behavior in insulators is very different from the oscillatory spatial dependence of the potential produced by a point charge in a free electron-like metal with Fermi momentum $k_F$, i.e. the Friedel oscillations which can lead one to incorrectly identify an absolute instead of relative attractive interaction between two charges at an average distance of $|\va{r}_1-\va{r}_2| = \frac{\pi}{2k_F}$ \cite{Friedel_original}. The crux of the issue is that the interaction between two particles in the presence of a polarizable medium is not given by the potential of one at the position of the other calculated without the other being present. If indeed the polarization clouds of the particles significantly overlap as they will at modest distances, there can and will be strong interference effects which could be constructive, as in the reduction of the on-site Coulomb repulsion between two electrons in atomic orbitals, or they could be destructive as in the enhancement of the Coulomb interaction in situations where two electrons are situated on atoms sandwiching a polarizable ion. The understanding of short-range Coulomb interactions is of special importance in solids for which the charge carriers propagate in relatively narrow bands, and in which the charge density and polarizability are strongly non-uniform, including the cuprate superconducting materials.  

Investigating this phenomenon requires the calculation of the response of a polarizable system in the presence of two charges separated by a distance $|\va{r}_1-\va{r}_2| = \sqrt{|\va{r}_1|^2 + |\va{r}_2|^2 - 2\va{r}_1 \cdot \va{r}_2}$. Here we place them at the positions of constituent ions with the approximation that the electric field produced by electrons in atomic orbitals can be well captured at distances beyond the radial extent of their wavefunction by fixing their charge at the nucleus. Corrections to this for oriented $3d$ or $4f$ orbitals rather than spherically symmetric $s$ orbitals are very small for distances between the charge and the polarizable atoms of the order of 2 \angstrom{} or greater. We also have to take into account that the net field on a polarizable atom is given by the sum of the point charge fields and of the fields of the surrounding induced dipoles. Quite generally, the polarizabilites of anions with orbitals of large radial extent are at least an order of magnitude larger than the ones of cations, so we consider only anion polarizabilities. In addition to the contribution of the polarizable atoms, we also consider the effective bond polarizabilities in order to take into account the covalency effects related to the hybridization between the valence electron wavefunctions of the cation ions with those of the negative anions. For example, in the $3d$ transition metal oxides, it is well established that these effects are rather large and that their properties are very sensitive to the energy difference and overlap between the cation and anion states \cite{nmr_cuprates, macridin_charge_density}. The introduction of point charges to such a system can locally modify those energy differences if the resultant fields at the cation and anion differ, generating a charge transfer-induced bond polarization that is highly directional and very different from the mainly isotropic atomic polarizability. Such a bond polarization would be accounted for in a band structure-based susceptibility calculations by including the appropriate interband transitions. 

In this work, we opt to employ a real space model to investigate the consequences of these polarization mechanisms. There are several reasons why we follow this path instead of using band theory momentum space methods. First, the effects we are discussing are particularly relevant for the systems in which short-range interactions are of extreme importance such as the $3d$ transition metal and rare earth compounds. Unfortunately, for such highly-correlated materials, band theory approaches often fail in capturing band gaps and the retention of local magnetic moments well above the ordering temperatures, especially in electron or hole-doped systems which exhibit strongly modified electronic structures \cite{lda_cuprate_correlated}. Secondly, the density functional theory (DFT) usually approaches employed are based on the use of single particle symmetry-restricted wavefunctions to describe the total ground state electronic density of the materials which is known to be a functional of the ground state energy. In DFT calculations of the susceptibility, one also uses these wavefunctions for the excited states up to very high energies in order to describe the response of a system to external perturbations, usually limited to the random phase approximation (RPA) which does not include local field effects (LFE) by default \cite{rpa_original_2, RPA_dft_mazin}. These local field corrections drive many of the effects we will discuss later and are extremely important to describe short-range interactions. Thirdly, the calculations of the effective interactions based on the sum of the screened potentials produced by one particle on the other can either underestimate or overestimate the results depending on the geometry of the overlapping induced polarization clouds. Of course, the drawbacks of the real space method we use is that we have to resort to model Hamiltonians which however involves parameters that mostly can be obtained from band structure calculations. 

In order to concretize our approach, including the associated mathematical physics relations and computational procedure, we analyze as a representative and telling example a square planar CuO$_2$ lattice, which is part of the cuprate high-temperature superconductors. The undoped low-energy electronic structure of these compounds consists of $S=\frac{1}{2}$ local magnetic moments on Cu$^{2+}$ ($3d^9$) ions with one hole in their $d_{x^2-y^2}$ orbital at the center of a square of oxygen ions in a formally closed shell O$^{2-}$ ($2p^6$) configuration \cite{zr_singlet}. The paper is structured as follows. We first describe in a general and material-agnostic way the theory characterizing the effects both atomic and bond polarizabilites, highlighting the differences between the two mechanisms and showcasing the nature of the interference between overlapping polarization clouds. Then, we detail our real space polarization model as applied to the the cuprate CuO$_2$ plane which is followed by the presentation and discussion of the resulting unusual two-particle screened non-monotonic Coulomb interaction. Finally, the importance of the physics extracted from these results is elaborated upon with respect to the high-temperature superconductor pairing puzzle, notably concerning their short measured Cooper pair coherence length and the inherent pairing anisotropy visible in the pseudogap preformed pair region of the $d$-wave superconductor phase diagram.

\section{Formation and Interference between Polarization Clouds}

\begin{figure}[h!]
	\includegraphics[width=0.9\textwidth]{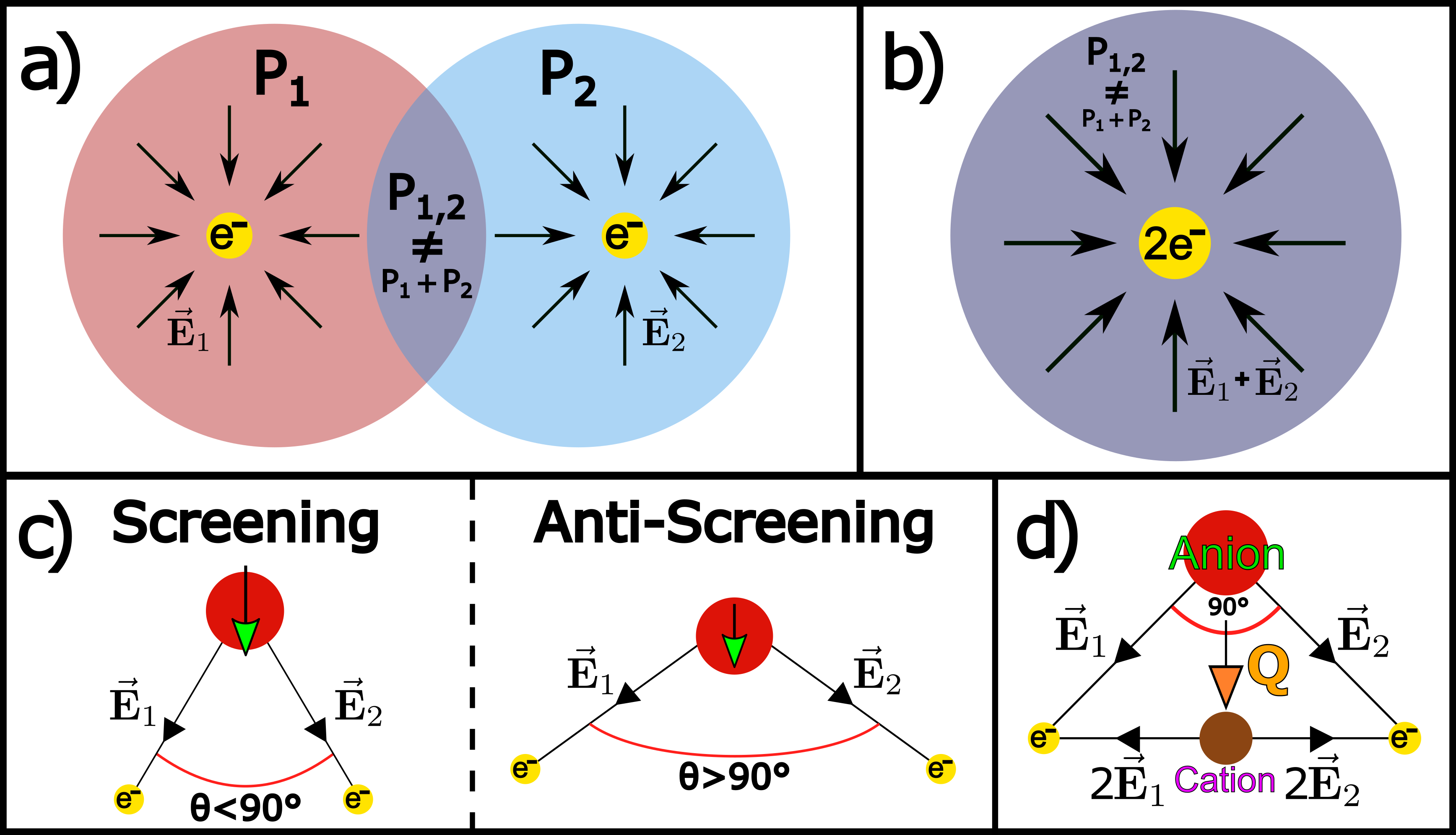}
	\caption{Interference of the polarization clouds of two charged particles. a) Diagram representing the nonlinear superposition of the polarization clouds $P_1$ and $P_2$ of two electrons. b) Complete overlap of the polarization clouds of two electrons on the same atomic site. c) Dependence of the screening of the Coulomb interaction between two electrons on the electron-ion-electron bond angle $\theta$ caused by their induction of atomic dipoles (green arrows) on the ion. d) Representation of the induction of an effective anion-cation bond dipole (orange arrow) due to an effective transfer of charge $Q$ induced by the electrons.}\label{fig:polarization_cloud_interference}
\end{figure}

When a point charge is introduced inside or near a material, say at position $\va{R}_{h_1}$, its electric field generates a polarization cloud in the form of electric dipoles induced in that material which are dependent on its optical (high energy) dielectric function. While such a polarization cloud has a complicated structure, it is rather local as depicted in Figure \ref{fig:polarization_cloud_interference} a)-b). It depends on the internal interactions of the induced charges described by the generalized dielectric function of the form $\epsilon(\va{r}_1, \va{r}_2)$, which in a random phase-like approximation (RPA) would be simplified to $\epsilon(|\va{r}_1-\va{r}_2|)$. The more general form includes the details of the interactions between the induced dipoles. When a second such charged particle is also introduced at a position $\va{R}_{h_2}$, it will similarly have a polarization cloud of induced dipole moments surrounding it. If the particles are far enough apart from each other such that their polarization clouds effectively do not overlap, then the potential felt by one particle is simply the potential generated by the other one screened by its individual polarization cloud. However, when the particles are closer, the situation becomes more complicated due to the non-trivial overlap of their polarization clouds. While the total monopole electric field at the positions inside of the overlapping region is the direct vector sum of the two particles' fields, the screened Coulomb potential $V'$ is dependent on the two-particle polarization energy cost $E^{(2)}_{pol}$ of introducing the polarizing external charges in the first place:
\begin{align}\label{eq:screened_potential}
	V'(\va{R}_{h_1}, \va{R}_{h_2}) = V_0(\va{R}_{h_1}, \va{R}_{h_2}) - E^{(2)}_{pol}(\va{R}_{h_1}, \va{R}_{h_2}),
\end{align}
\begin{align}\label{eq:epol_subtraction}
	E^{(2)}_{pol}(\va{R}_{h_1}, \va{R}_{h_2}) = E_{pol}(\va{R}_{h_1}, \va{R}_{h_2}) - E_{pol}(\va{R}_{h_1}) - E_{pol}(\va{R}_{h_2})
\end{align}
where $V_0$ is the bare Coulomb interaction and the superscript of $E^{(2)}_{pol}$ refers to the fact that the single particle contributions to polarization are subtracted in order to determine the two-particle screening. This polarization energy, based on the tendency of dipole moments $\va{p}$ to align with electric fields $\va{E}$, depends on terms proportional to $\va{p} \cdot \va{E}$ summed over the entire polarization clouds. Under linear response theory, since dipole moments $\va{p}$ are themselves directly proportional to electric fields $\va{E}$ and to atomic polarizabilities $\alpha$, $E_{pol}$ (including single particle terms) is directly dependent on the dot product of the perturbing electric fields which for the simple example of two charges with fields $\va{E}_1$ and $\va{E}_2$ has the form:
\begin{align}\label{eq:pol_energy_initial_definition}
	E_{pol} \propto \frac{1}{2} \va{p} \cdot \va{E} = \frac{1}{2}(\alpha \va{E}^2 ) = \frac{1}{2}(\alpha |\va{E}_1 + \va{E}_2|^2 ) = \frac{\alpha}{2}\left( |\va{E}_1|^2 + |\va{E}_{2}|^2 + 2\va{E}_1 \cdot \va{E}_2  \right).
\end{align}

This relation can lead to drastically different polarization outcomes based on the nonlinear combined effect of multiple polarization clouds, as diagrammatized in Figure \ref{fig:polarization_cloud_interference} a)-b). A very important problem regarding the interaction between two equal point charges at short distances is that their polarization clouds can strongly overlap, such that their interaction is not described by the sum of the individual screened potential generated by one at the position of the other. Indeed, as depicted in Figure \ref{fig:polarization_cloud_interference} a), local field effects can strongly modify the nature of polarization clouds, especially in the regions of high overlap. To further illustrate this concept, consider the scenario shown in Figure \ref{fig:polarization_cloud_interference} b) in which two charges are occupying the same atom, leading to a complete overlap between their polarization clouds. In that case, based on Equation \eqref{eq:pol_energy_initial_definition}, the bare Coulomb onsite interaction $U_0$ is screened twice as strongly as the result obtained through the simple approach of linearly adding the screening effect of the two charges due to the $2\va{E}_1 \cdot \va{E}_2$ term, i.e. the two-particle part $E^{(2)}_{pol}$.

A direct consequence of this is a potentially strong modulation of short-range screened Coulomb interactions based on material geometry, which reinforces the need to go beyond the Clausius-Mossotti approach in order to capture the LFE that govern the interactions at this range. A simple example of the importance of the non-trivial interference of polarization clouds is showcased in Figure \ref{fig:polarization_cloud_interference} c). Since the deviation of the bare Coulomb interaction between two nearest neighbor (or sufficiently close) charged particles with polarizable ions placed between them directly depends on $\va{E}_1 \cdot \va{E}_2 = |\va{E}_1||\va{E}_2|cos(\theta)$, bond angles have a strong influence on both screening strength and whether screening or anti-screening occurs. Based on Equation \eqref{eq:screened_potential}, if $\theta$ is larger than 90\degree{}, the induced dipoles anti-screen the two-particle interaction leading to an increase in repulsion if their charge has the same sign. However, if that angle is smaller than 90\degree{}, the opposite occurs and the Coulomb repulsion is weakened. There has been interest in superconductors with similar geometries since such a reduction in fermionic electrostatic repulsion can represent a significant piece of the pairing interaction puzzle in high-temperature superconductors \cite{FeAs_Sawatzky, sawatzky_pnictides}.

On the other hand, bond dipoles can cause screening even in the $\theta = 90$\degree{} geometry contrary to the atomic dipole screening and anti-screening mechanisms. As showcased in Figure \ref{fig:polarization_cloud_interference} d), if the electronic states of the polarizable anion have covalency with some cation, the potentials generated by the charged particles have different magnitudes between the anion and cation locations. This induces a modification of the charge transfer energy $\Delta = \Delta^0 + |e|(V_{anion} - V_{cation})$ between these two, leading to an effective charge transfer to attain a new electrostatic ground state. This is precisely this exchange of charge that causes the formation of bond dipoles, which by their nature are highly anisotropic because their orientation is constrained along the anion-cation bond axis. Of course, from a modern theory of polarization point of view, the polarizing effect of the charged particles leads to a deformation of the shared electronic orbital in the bond which results in a displacement of the effective Wannier centers \cite{modern_theory_of_polarization_precursor, modern_theory_of_polarization_original}. As previously mentioned however, we consider the transfer of point charges between atomic nuclei to take this phenomenon into account in our model, which will be sufficient to exhibit its physical consequences.

\section{C\MakeLowercase{u}O$_2$ Polarization Model}

In order to investigate the impact of polarization anisotropy, LFE and the interference effects of two distinct polarization clouds, we have chosen to focus our attention on the two-dimensional CuO$_2$ system, which is known to host the important physics regarding superconductivity in the cuprates. We note that the cuprates are part of the perovskite family of crystal structures, which is common for a large number of compounds which garner modern attention because of their wide diversity of properties including superconductivity.  We start with the conclusion of most studies that the mobile charge carriers are primarily housed in O $2p$ orbitals in the hole-doped cuprates, as in the Zhang-Rice singlet description \cite{zr_singlet} or the three-spin polaron picture of Emery \cite{emery_cuprate}, because it is widely recognized that the parent compounds are in the charge transfer gap region of the ZSA classification scheme \cite{zsa_original_paper, sawatzky_hubbard_u}. In this work, we are interested in the screened Coulomb interaction $V'(\va{R}_{h_1}, \va{R}_{h_2})$ from Equation \eqref{eq:screened_potential} between two doped holes $h_1$ and $h_2$ of charge $\abs{e}$ located on the ions positioned at $\va{R}_{h_1}$ and $\va{R}_{h_2}$ respectively in an initially undoped CuO$_2$ layer with a typical Cu-Cu distance $a = 3.80 \; \angstrom$ \cite{cuprate_dft_params}.

\begin{figure}[h!]
	\begin{center}
		\includegraphics[width=\textwidth]{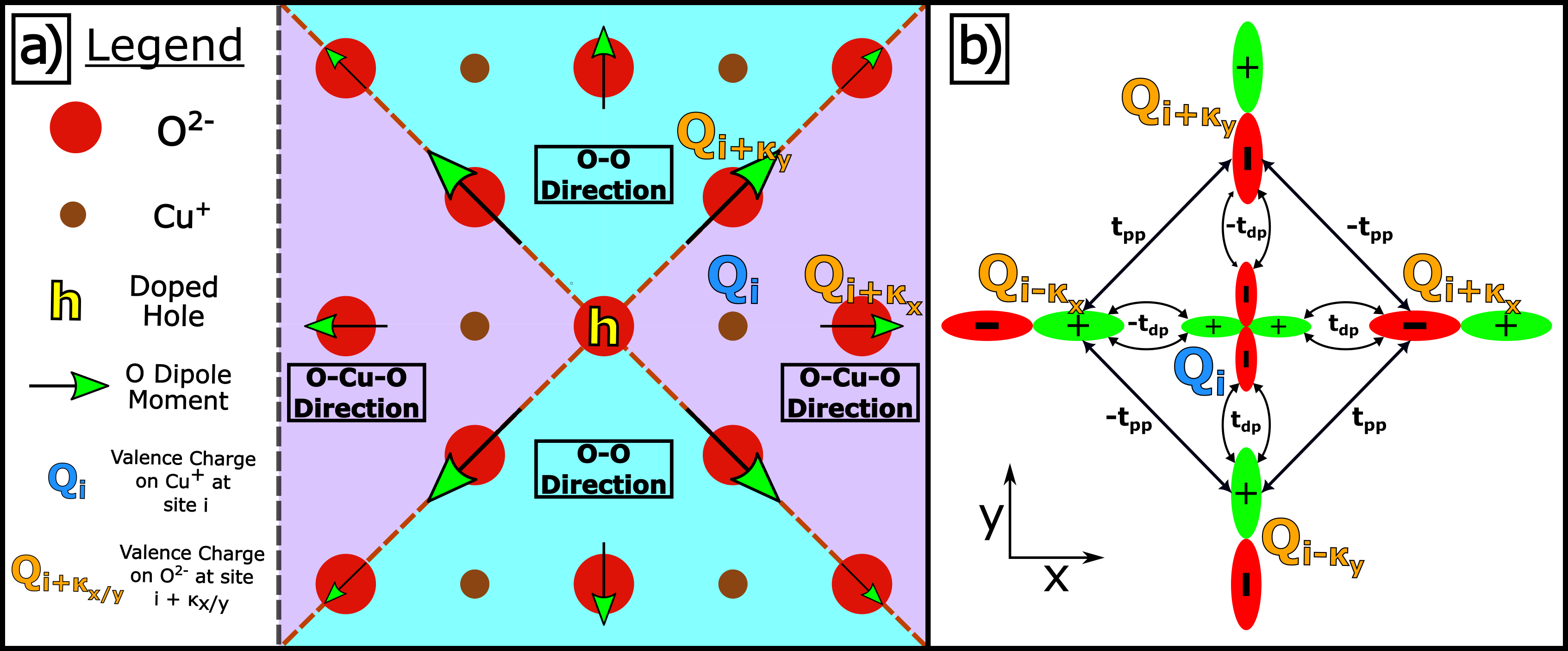}
		\caption{a) Diagram of the polarization effects induced by a single doped hole $h$ on an O site of a CuO$_2$ cluster. b) Nearest neighbor hole hopping processes between a single Cu $3d_{x^2-y^2}$ orbital and its four surrounding O $2p_x$/$2p_y$ orbitals. This is represented by the Hamiltonian in Equation \eqref{eq:CuO_Hamiltonian}.}\label{fig:model_diagram}
	\end{center}
\end{figure}

We set up an electrostatic model illustrated in Figure \ref{fig:model_diagram} a). We found that a finite square cluster large enough to include a square of at least size 5$a$ $\times$ 5$a$ surrounding each doped hole is sufficient for convergence due to the diminishment of LFE at large distances.  Each standard unit cell containing one Cu ion and two O ions is labeled by a two-dimensional index $i$, with the O positions indicated by $i + \kappa_{x/y}$ where $\kappa_{x/y} = \frac{a}{2} \vu{x}/\vu{y}$. Atomic dipoles centered on ionic positions are represented by $\pion$, and the effective valence charge of an ion (which is used to calculate bond dipoles) is $\Qion$. To compute the screened Coulomb interaction $V'(\va{R}_{h_1}, \va{R}_{h_2})$, we fix $h_1$ on a specific ion and vary the position of $h_2$. Crucially, when considering doped holes on O sites, there is an angular dependence to the two-hole interaction caused by the presence of twice as many O ions as Cu ions in the structure, leading to a two-fold rotation-symmetric Coulomb interaction $V'(\va{R}_{h_1}, \va{R}_{h_2}) \neq V'(\Rh)$ that is not a pure function of the hole-hole distance $\Rh = R = |\va{R}_{h_2} - \va{R}_{h_1}|$. This is in contrast with the Cu-centered perspective since a four-fold rotational symmetry exists instead in that case. Consequently, we define two different orientations (with respect to the position of the first hole) in which we can introduce a second hole: the ``O-Cu-O direction'' which features a Cu ion in between next nearest neighbor O sites, and the ``O-O direction'' which does not. As shown in the presentation of our results further below, these two orientations showcase a pronounced and physically-meaningful angular dependence of the screened Coulomb interaction which has ramifications for superconductivity.

The two polarization mechanisms we take into account, namely atomic dipoles and bond dipoles, have different physical origins and must be implemented separately. These depend on the total electric field $\va{E}_i$ and electric potential $V_i$ at ionic positions, which include both monopole and dipole contributions. A more detailed breakdown of all of these parts is presented in the Supplemental Material (SM) \cite{supp_material_cuprate} (see also references \cite{O_ionization_potential, O_electron_affinity, O_second_electron_affinity, SciPy, charge_transfer_madelung_vacuum, cuprate_experimental_delta_1.5, octopus_local_field, cuprate_phase_diagram} therein).

\subsection{Atomic Dipoles}

This model's handling of atomic dipoles is simple and straightforward. They are treated as point dipoles centered at the nucleus of each polarizable anion. Their magnitude and direction is directly proportional to the total electric field from all other sources in the real space finite cluster, which includes the doped holes, the bond dipoles and the other atomic dipoles:
\begin{align}\label{eq:O_dipoles}
	\pion = \alpha_i \va{E}_i,
\end{align}
\noindent where $\alpha_i$ is the atomic polarizability of the ion at position $i$. In the case of CuO$_2$, we have found that Cu cation dipoles are inconsequential due to their very small polarizability; we thus focus on O dipoles with $\alpha^0_O$ = 2.75 \angstrom{}$^3$ as a literature baseline parameter \cite{O_polarizability}.

\subsection{Bond Dipoles}

The bond polarization mechanism is taken into account by calculating the effective charges transferred between each Cu-O bond due to the introduction of the polarizing doped holes to the system. In other words, we only consider the difference in the ionization charge at each site with respect to the charges in the undoped system. We start with the hole vacuum state populated by O $2p^6$ and Cu $3d^{10}$ configurations such that we only have filled electron shells \cite{zr_singlet, emery_cuprate, apres_cuprate_electronic_structure}. It is known experimentally as well as through \textit{ab initio} calculations that the CuO$_2$ planes in undoped cuprates host 1 hole per Cu \cite{cuprate_dft_params, neutron_scattering_afm}, leading to the O $2p^6$ and Cu $3d^9$ configurations. While that hole is often taken as being fully on the Cu $3d$ orbitals, calculations and nuclear magnetic resonance (NMR) measurements have shown a strong covalent character in the wavefunction of that hole; approximately 70\% to 80\% of its charge density rests on Cu sites, while 20\% to 30\% is on O ions (corresponding to 10\% to 15\% per O) \cite{eskes_delta, macridin_charge_density, nmr_cuprates, cuprate_nmr_hole_density}. We use this experimentally-determined degree of covalency as way to calibrate our model by assigning a baseline undoped Cu-O charge transfer energy $\Delta^0$ = 6.0 eV (see the SM) \cite{supp_material_cuprate}.

To compute the valence charges we need to determine $\Delta_{i,\kappa}$, the local charge transfer energy between bonded ions at positions $i$ and $i+\kappa$. This quantity is defined as the difference between the energy costs $\mathcal{E}_{i}$ of adding a hole to the ion at $i+\kappa$ and to the ion at $i$: 
\begin{equation}\label{eq:charge_transfer}
	\Delta_{i,\kappa} = \mathcal{E}_{i+\kappa} - \mathcal{E}_{i},
\end{equation}
\begin{align}\label{eq:on-site_energy}
	\mathcal{E}_{i} = |e|V_i.
\end{align}
\noindent Depending on the placement of the two doped holes, the charge transfer energy of each bond will be uniquely affected by polarization effects. For CuO$_2$, we are interested by nearest neighbor Cu-O bonds.  

Then, a local single-particle Hamiltonian $H_{i}$ is set up for each Cu ion and its four surrounding O ions (``CuO$_4$'' clusters). $H_{i}$ includes the modulated charge transfer energy $\Delta_{i, \kappa}$ for each of its four bonds, as well as the nearest neighbor O-Cu-O and O-O hopping channels as shown in Figure \ref{fig:model_diagram} b), whose strengths are respectively controlled by hopping integrals $t_{dp}$ and $t_{pp}$:
\begin{equation}\label{eq:CuO_Hamiltonian}
	H_{i} = \mqty[0 & -t_{dp} & - t_{dp} & t_{dp} & t_{dp} \\ -t_{dp} & \Delta_{i,\kappa_y} & t_{pp} & 0 & -t_{pp} \\ -t_{dp} & t_{pp} & \Delta_{i,-\kappa_x} & -t_{pp} & 0 \\ t_{dp} & 0 & -t_{pp} & \Delta_{i,-\kappa_y} & t_{pp} \\ t_{dp} & -t_{pp} & 0 & t_{pp} & \Delta_{i,\kappa_x}].
\end{equation}
\noindent We set $t^0_{dp}$ = 1.30 eV and $t^0_{pp}$ = 0.65 eV as baseline values based on literature results obtained from experimental data and tight binding fits to DFT-calculated band structures on cuprates \cite{cuprate_review_params, cuprate_dft_params}.

Since there is effectively one hole per Cu ion that is shared with their neighboring O in undoped cuprate CuO$_2$ planes, diagonalization of $H_{i}$ allows us to determine the distribution of this charge between these ions. In other words, diagonalizing $H_{i}$ in Equation \eqref{eq:CuO_Hamiltonian} for all the CuO$_{4}$ clusters included in our model yields their ground states from which we extract the effective charge on each ion. From there, the potentials and fields originating from bond dipoles is simply the combination of the monopole potentials and fields from all induced charges.

\subsection{Polarization Energy}

To calculate $E^{(2)}_{pol}(\va{R}_{h_1}, \va{R}_{h_2})$ from Equations \eqref{eq:screened_potential}-\eqref{eq:epol_subtraction} for the effective two-hole screened interaction, subtraction of single hole contributions and of the base undoped energy of each cluster is necessary. In this way, the actual two-particle screened Coulomb interaction can be isolated from the single-particle part of the Coulomb energy. Let us denote a hole configuration with $h$ which can indicate the presence of two holes ($h = h_1 + h_2$), a single hole ($h = h_1$ or $h = h_2$) or no holes ($h = 0$) such that:
\begin{align}\label{eq:energy_configuration}
	E^{h}_{pol}(\va{R}_{h_1}, \va{R}_{h_2}) = \sum_i \left[ \frac{\pion}{2}  \cdot \va{E}^{mono}_i - Q_i V^{mono}_{i} \right]
\end{align}
\begin{align}\label{eq:energy_subtraction}
	E^{(2)}_{pol}(\va{R}_{h_1}, \va{R}_{h_2}) &= (E^{h_1+h_2}_{pol} - E^0_{pol}) - (E^{h_1}_{pol} - E^0_{pol}) - (E^{h_2}_{pol} - E^0_{pol}) \nonumber \\ &= E^{h_1+h_2}_{pol} - E^{h_1}_{pol} - E^{h_2}_{pol} + E^0_{pol},
\end{align}
\noindent where $\va{E}^{mono}_i$ and $V^{mono}_i$ are the monopole contributions to the electric field and potential. Equation \eqref{eq:energy_configuration} includes monopole-monopole, monopole-dipole, dipole formation energy and dipole-dipole interactions \cite{exact_calcs_local_field}, the latter being the originator of local field effects. The sum is over all ions, so both Cu and O sites are included in this case, and double counting of monopole-monopole interactions are taken into account (see the SM) \cite{supp_material_cuprate}.

\section{Polarization Results and Discussion}

\subsection{Screened Coulomb Interaction}

\begin{figure}[h!]
	\includegraphics[width=0.9\columnwidth]{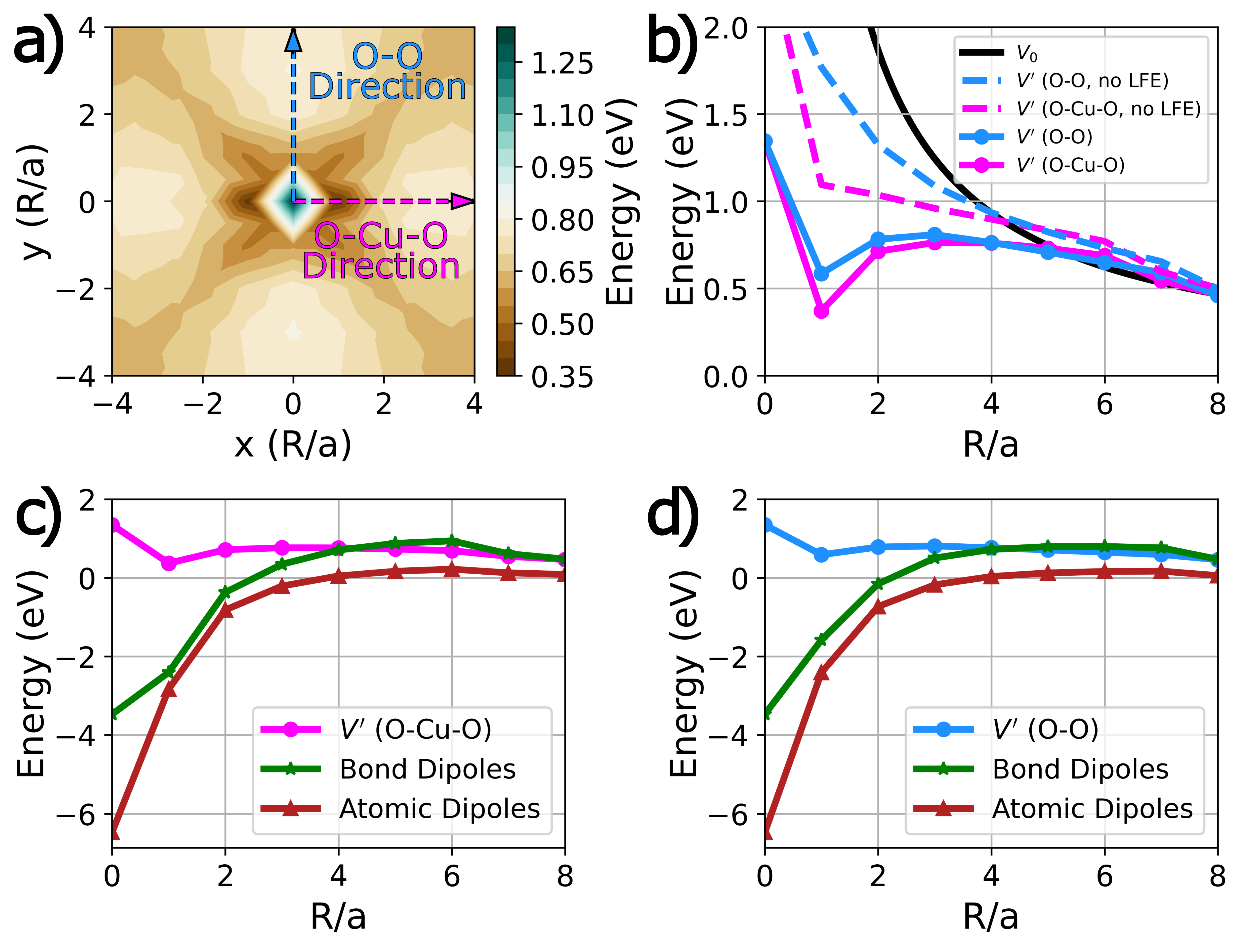}
	\caption{Screened two-hole Coulomb interaction $V'(\va{R}_{h_1}, \va{R}_{h_2})$ with literature base Hamiltonian parameters (see text) and $h_1$ fixed at $R=0$ on an O site as a function of the position of $h_2$ on other O sites. a) Real space Coulomb interaction color plot with $h_1$ at the center following the geometry of Figure \ref{fig:model_diagram} a). b) Comparison of $V'(\va{R}_{h_1}, \va{R}_{h_2})$ with $h_2$ placed along the O-Cu-O versus the O-O direction. Curves excluding LFE are also shown. A breakdown of the different contribution to $E^{(2)}_{pol}(\va{R}_{h_1}, \va{R}_{h_2})$ from Equation \eqref{eq:energy_subtraction} is shown in the c) O-Cu-O and d) O-O directions.}\label{fig:base_potential}
\end{figure}

The screened Coulomb interaction between two holes on O atoms calculated by solving the nonlinear system of equations formed by Equation \eqref{eq:O_dipoles} and the diagonalization of Equation \eqref{eq:CuO_Hamiltonian}, with one of them at the origin and the other at various distances and directions is shown in Figure \ref{fig:base_potential}. In subfigure a), we depict it in a contour plot in which the interaction strength is indicated by the color scale on the right. We see a large repulsion of about 1.5 eV when both holes are on the same O at the origin. We note that over a range of distances between the holes of up to 5 lattice constants (about 15 $\angstrom{}$), excluding the same-site value, the screened Coulomb interaction varies by less than 1 eV and contains minima and maxima at similar distances but different directions. The maximum reduction occurs when the holes are separated by a lattice constant along the O-Cu-O direction, i.e. when a Cu ion lies between two hole-occupied next nearest neighbor O ions. This actually represents a local minimum in the interaction. A somewhat smaller ``minimum'' is also found along the O-O direction. This effect is a direct result of the dominating contribution of the Cu-O bond polarizability to the directionality of the screening which, as noted above and shown in Figure \ref{fig:polarization_cloud_interference}, has a tensorial character. It is also very interesting that there are strips of much weaker interaction minima along the diagonal directions. This is a fascinating landscape of interactions with consequences for potential charge density waves both along the diagonal directions, and along the Cu-O bond directions for larger doping concentrations. It also represents a low repulsive energy contribution for pairing, and eventually superconductivity with relatively short coherence lengths and enhanced effective attractive interactions via the exchange of bosons. 

In Figure \ref{fig:base_potential} b), the interactions specifically along the O-O and O-Cu-O directions are juxtaposed as solid lines. The results of the same calculations without considering local field effects are also presented through dashed lines. The strong contributions of the local field effects at short distances (up to about 5 lattice constants) are clear, especially for the nearest and next nearest neighbor O distances. Furthermore, the important physical difference between the two directions is again flagrant. Indeed, the screening influence of the Cu-O bond polarizability is key to explaining the appearance and angular dependence of this repulsion minimum. As visualized in Figure \ref{fig:model_diagram} a), the four oxygen ions surrounding a copper site form a square and thus the O-O-O angle is 90\degree{}. Consequently, as shown in Figure \ref{fig:polarization_cloud_interference}, this results in a vanishing contribution to the polarization energy from ionic polarizability. The local minima bring the Coulomb interaction down to values well below 0.5 eV which, as shown below, can be lowered further with some changes in the model's parameters. We also note the nearly constant interaction at distances between 2 and 5 lattice constants spanning a modest local maximum. Moreover, the atomic and bond polarizability contributions to the polarization energy are presented in Figure \ref{fig:base_potential} b) and c). This data shows that these two parts have similar magnitudes but distinctly different behavior as a function of distance. The strong non-monotonicity of $V'(\va{R}_{h_1}, \va{R}_{h_2})$ and the deeper repulsion minimum at $R=a$ in the O-Cu-O direction is demonstrated to originate from the bond dipole contribution to screening.

\begin{figure}[h!]
	\includegraphics[width=0.9\columnwidth]{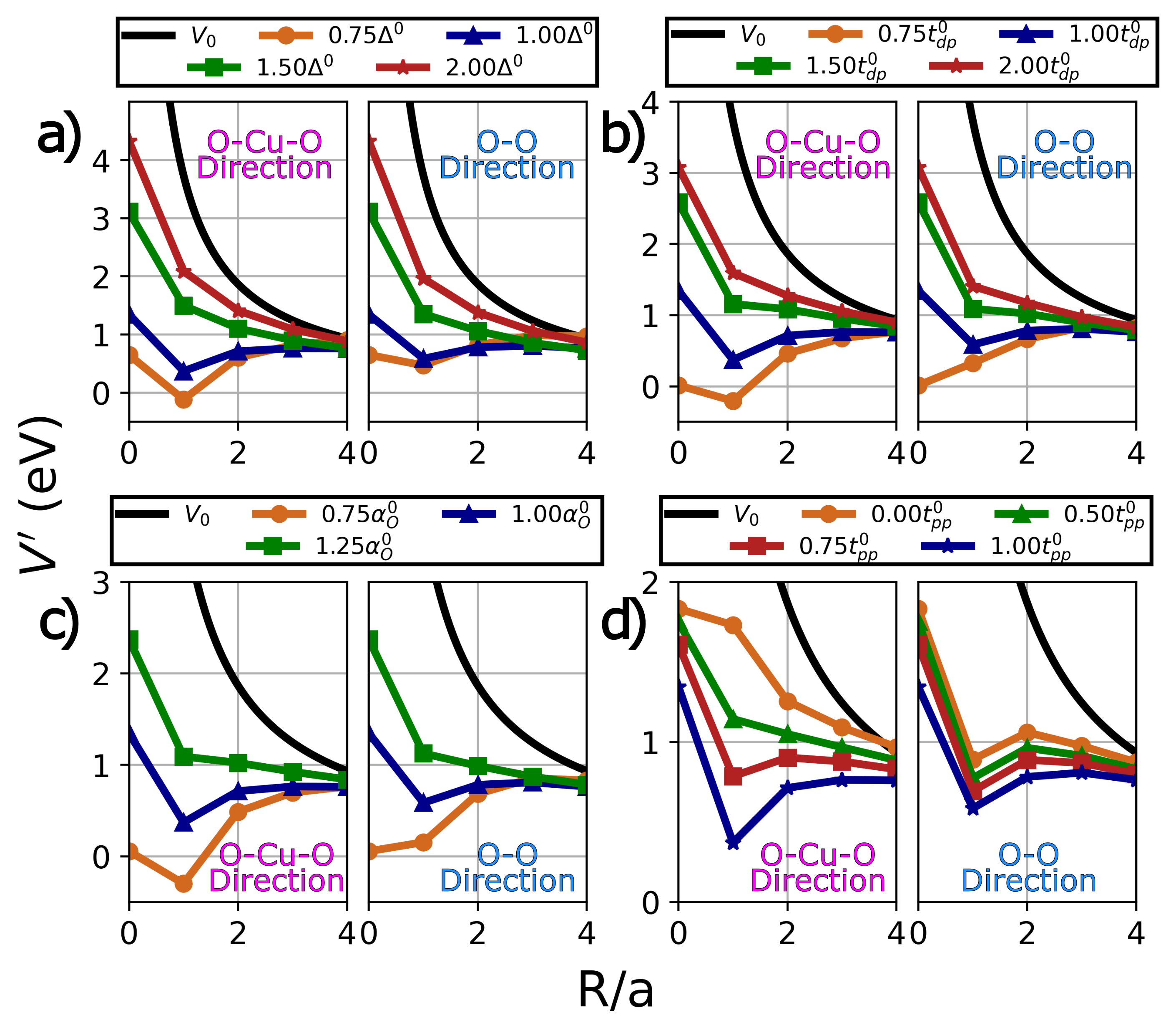}
	\caption{$V'(\va{R}_{h_1}, \va{R}_{h_2})$ calculated for various Hamiltonian parameter values. With the two holes on O sites (and one fixed at $R=0$), potentials with different a) $\Delta^0$, b) $t_{dp}$, c) $\alpha_O$ and d) $t_{pp}$ values are shown while keeping all other parameters equal to their base values ($\Delta^0$ = 6.0 eV, $t^0_{dp}$ = 1.30 eV, $\alpha^0_O$ = 2.75 $\angstrom{}^3$ and $t^0_{pp}$ = 0.65 eV).}
	\label{fig:parameters_potential}
\end{figure}

We also studied the dependence of $V'(\va{R}_{h_1}, \va{R}_{h_2})$ on the model's parameters as shown in Figure \ref{fig:parameters_potential}. Through a), b) and c) we see that the screening strength is inversely proportional to $\Delta^0$, $t_{dp}$ and $\alpha_O$, which makes sense since large values mean that polarization effects will not modulate covalency and atomic dipole magnitudes as strongly relative to the undoped state as illustrated by Equations \eqref{eq:charge_transfer} and \eqref{eq:on-site_energy}. On the other hand, reducing it leads to the O-Cu-O direction $\Rh=a$ minimum to be attractive, indicating that strong deviations from literature parameters can lead to unphysical results (such as $\Delta_{i,\kappa} < 0$ for some Cu-O bonds) and convergence issues, corroborating the model's realism. Thus, similar calculations for other materials necessitate somewhat accurate approximations to Hamiltonian parameters. Moreover, d) reveals that lowering the nearest neighbor O-O hopping has the peculiar effect of destroying the O-Cu-O direction local minimum and even turn it into a maximum. Physically, the O-O hopping channel dying out means that more Cu-O charge transfer occurs, causing a large buildup of positive charge on the central Cu which repels the holes and elevates the monopole and same-site parts of $V'(\va{R}_{h_1}, \va{R}_{h_2})$. These results are consistent with the experimental trend of cuprates with stronger oxygen-oxygen hopping strengths (particularly the next nearest neighbor hopping $t'_{pp}$), larger in-plane O-Cu-O distances leading to weaker Cu-O hopping, and smaller Cu-O charge transfer energies to exhibit higher $T_{C}$ values \cite{cuprate_tpp_tc_1, cuprate_tdp_tc, cuprate_tpp_tc_2}. It is interesting to note that there is no anisotropy in the Cu-Cu interaction, but there is an appreciable difference between the O-Cu-O and O-O directions in the O-Cu-O interaction. Additionally, Figure \ref{fig:parameters_potential} shows the absence of long range screening characteristic to 1D and 2D materials since a long-range dielectric constant cannot be defined for dimensions lower than three, but on the other hand low-dimensional systems usually exhibit stronger local field effects \cite{low_dim_no_screening_1, low_dim_no_screening_2, 2d_clausius_dynamical}. While apical oxygen sites were excluded from this model, their inclusion would keep the system quasi two-dimensional and thus conserve this long range behavior. Moreover, we note that including apical oxygen ions in these calculations did not appreciably modify our results; a slight constant screening increase explained by the generation of additional atomic dipole contributions in the style of Figure \ref{fig:polarization_cloud_interference} c) was observed. These were also smaller due to the larger apical Cu-O distance compared to the in-plane one in most cuprates. While we opted to exclude them in this model since they are not necessary to capture the interesting screening structure we discuss, further calculations investigating interactions along the inter-plane direction in more details are in the cards, especially since those have been shown to have an impact on in-plane exchange interactions in the context of superconductivity \cite{cuprate_apical_oxygen}. An interesting extension of this analysis would be the investigation of the two-particle interaction when they are both placed on Cu sites or one on each ion type instead of having both of them of O ions. However, due to the high Cu on-site interaction, doped holes occupying $d$ states is energetically costly and thus one should also analyze the electron-doped cuprate systems.

\subsection{Pairing in High-temperature Superconductors}

Why is a local minimum in the pair coulomb interaction important in obtaining high-temperature superconductors? It is generally accepted that the superconductivity is the result of a net effective attractive interaction between two fermions of opposite spin, which in BCS theory is driven by the exchange of phonons or in general bosons between them. However, this competes with the (screened) direct Coulomb repulsion. It is also accepted that the effective average distance between the two fermions in the superconducting state is given by the coherence length $\epsilon_{BCS} = \frac{\hbar v_F}{\Delta_{gap}}$ where $v_F$ is the Fermi velocity and $\Delta_{gap}$ is the superconducting gap which grows with increasing T$_C$ \cite{coherence_length}. However, in conventional descriptions, the screened Coulomb repulsion increases with interparticle distance as $\frac{1}{R}$ or even exponentially, which normally would require an even stronger growth of the electron-boson interaction to obtain a high T$_C$. Intuitively, this competition generally would favor rather long coherence lengths for pairing. Nonetheless, we demonstrated that there are strong minima in the screened Coulomb interaction at short distances and a large region with an almost constant Coulomb repulsion at a distance between 2 and 5 lattice constants. This kind of behavior could promote short coherence lengths and T$_C$’s much higher than those one might expect if these minima and regions of constant repulsion were not present. Indeed, this high energy electron-hole polarization-induced screening fulfills the obligation for a strong reduction in Coulomb repulsion for Cooper pairing. This lowers the need for retardation effects in low energy scale boson-driven superconductors, allowing for larger boson exchange effects since this reduction occurs at a short-range permitting the paired electrons to be closer together compared to BCS superconductors. It must be made clear that the polarization-induced local repulsion is not alone directly responsible for the short coherence length of superconducting pairing; it supports the possibility of an attractive interaction at distances too short for conventional BCS superconductivity to emerge. Of course as usual in complicated situations, ``the devil is in the detail''. We nevertheless point out that the assumption of a Thomas-Fermi screened potential or the Debye-like screening in semiconductors is very strongly modified by the local field corrections and the interference of overlapping screening clouds of electrons in solids which can in fact favor rather short-range pairing of the order of lattice parameters. This effect however critically depends on the spatial location of the charges i.e. it is a strong function of $(r,r’)$ rather than one given by only $|r-r’|$ as assumed in approximating the dielectric function in momentum space as a function a single wavevector $\epsilon(q)$, and directly emerges from the interference of overlapping screening clouds of distinct particles. In addition to the high energy effects detailed in this work, screening from the electron gas itself plays a part, but the high mass of the quasiparticles and and their incoherent nature away from the Fermi energy make them inefficient for screening.

Additionally, these results are consistent with the cuprates being $d$-wave superconductors and the anisotropy inherent to the experimentally-measured pseudogap phase considered to be intrinsically linked to the superconducting phase of these materials, which exhibits insulating states allowing for Cooper pair formation only along the O-Cu-O bond direction \cite{varma_quantum_flucts, varma_pseudogap, varma_pseudogap_experimental, pseudogap_cuprate_specific_heat, pseudogap_cuprate_nmr}. Phases characterized by similar pseudogaps have also been observed in other high-T$_C$ superconductors such as pnictides and nickelates \cite{pseudogap_pnictides, pseudogap_nickelates}, but also in conventional superconductors \cite{pseudogap_Al, pseudogap_NbN, pseudogap_TiN} and non-superconducting materials \cite{pseudogap_manganite}, highlighting the broad applicability of this model. In principle, this polarization-induced anisotropy could potentially be measured through resonant inelastic X-ray scattering, which allows one to extract a two-electron (or hole) correlation function \cite{rixs_correlation_function}. Linear dichroism between signals generated from incident photons linearly polarized along the x versus the y axis as defined in Figure \ref{fig:model_diagram} could distinguish the hole-hole Coulomb interaction screening strengths along the O-Cu-O and O-O directions, which would support the experimentally-measured two-fold polarization rotational symmetry of cuprates \cite{cuprate_linear_dichroism}. Doped holes on O sites can either be surrounded by two Cu ions along the x axis or along the y axis, leading to $p_x$/$p_y$ orbital dichroism depending on the distribution of these holes. Specifically, YBCO is a promising candidate for such experimental exploration because of the presence of CuO chains in its structure which can further break $p_x$/$p_y$ degeneracy \cite{YBCO_chains}.

It must be clarified that while the anisotropy of interactions in cuprate planes has been discussed in the literature, such as through the charge density wave-mediated interaction between Fermi liquid quasiparticles \cite{Caprara_cuprate_cdw_anisotropy} which is anisotropic because of its striped nature, we present here a distinct and more fundamental manifestation of the inherent anisotropy of the polarization of this system through direct calculations of the screened Coulomb interaction between two charges centered on O sites. Furthermore, attractive two-particle interactions have been found in cuprates and even in the non-superconducting material SrVO$_3$ through real space RPA screening calculations based on LDA DFT calculations \cite{cuprate_U_W_real_space}. However, our results vary from these because the minima that have been found in that article are all within the interatomic O-Cu distance, a length scale even shorter than the ones we discuss. Investigation of interatomic charge placements is interesting and a natural extension to our current model. A subtraction procedure also had to be implemented in order to remove the metallic screening originating from the erroneous identification of undoped cuprates as metals that LDA DFT is guilty of. Moreover, the emergence of minima in SrVO$_3$ reinforces the vast applicability of our model; we want to reiterate that the investigation of polarization in cuprates and its implications for high-temperature superconductivity in this article is but a representative example of important polarization phenomena that strongly impact correlated materials, not only superconductors.

One might question how to incorporate our effective interactions in an Eliashberg-like calculation of T$_C$ \cite{eliashberg_original, dynes_original}. While this obviously is not a simple question, the strong deviation at short and large distances from the standard decreasing monotonic Coulomb repulsion we obtained disqualifies the separation of the average attractive interaction represented by $\lambda$ in the Allen-Dynes formulation from the average Coulomb repulsion represented by $\mu^*$ for relevant distances if indeed the coherence length is less than about 2-3 nm. The least one could do for predictions using the standard Allen-Dynes or similar relations employing a $\mu^*$ in its usual 0.1 to 0.15 range is to use the BCS relations and band structure calculations to determine a coherence length. If it is less than 2-3 nm, one could use our proposed simplified classical polarizability model to check for possible strong deviations from conventional distance-dependent screened interactions when including both local field corrections and the overlap interference effects of the charges' polarization clouds. All of these are taken into account in our real space approach. For this purpose, we would suggest to use only the high energy part of the polarizability in the screening, which is instantaneous.

\section{Conclusion}

Our model and results pave the way for proper, real space treatment of the screened fermionic Coulomb interaction in other materials, especially unconventional superconductors that feature highly polarizable ions and non-trivial covalency. The non-trivial interference of the polarization clouds generated by multiple doped charges leads to a pronounced anisotropic screening of their Coulomb repulsion at short-range, strengthening attractive bosonic exchange and diminishing the need for retardation effects, explaining the short coherence lengths of high temperature superconductors. Systems with measured pseudogap-like phases such as the iron pnictides or nickelates are good candidates to probe the important impact of local field effects. A further extension to this study is to analyze the dependence of T$_C$ on hole doping concentration, similarly to the doping-dependent paraelectric phase decline in SrTiO$_3$ which has been recently studied \cite{polarization_length_scale_doping}.

\begin{acknowledgments}
	This research was undertaken thanks in part to funding from the Max Planck-UBC-UTokyo Center for Quantum Materials and the Canada First Research Excellence Fund, Quantum Materials and Future Technologies Program, as well as by the Natural Sciences and Engineering Research Council (NSERC) for Canada.
\end{acknowledgments}

%
\end{bibunit}

\begin{bibunit}
\newpage
\subsection{Supplemental Material: Polarization Model Details}

\renewcommand{\theequation}{S.\arabic{equation}}
\renewcommand{\thefigure}{S.\arabic{figure}}

\subsection{System of Equations for Polarization Variables}

Our polarization model involves numerically solving a nonlinear system of equations for polarization variables, namely the atomic polarizabilities $\pion$ (defined by Equation (\textcolor{red}{4}) in the main text)  and the valence charges $\Qion$ (determined by the diagonalization of Equation (\textcolor{red}{7})) for all charges. These quantities are interdependent in the sense that the modification of the valence charges due to local charge transfer energy modulation have their own electric fields and potentials, which influence the atomic dipoles and vice versa. 

The total electric field $\va{E}_i$ and potential $V_i$ at a position $i$ can be broken down in terms of components with different physical origins, which helps to highlight the physical relationship between the polarization variables. There are the monopole contributions $\va{E}_{h,i}$ and $V_{h,i}$ from the doped holes themselves, the dipole contributions $\va{E}_{p,i}$ and $V_{p,i}$ from the atomic dipoles, and the contributions from the induced valences charges $\va{E}_{Q,i}$ and $V_{Q,i}$ which represent the bond dipole phenomenon. We have used the following explicit forms of these fields and potentials (including the bare potential interaction $V_0(\va{R}_{h_1}, \va{R}_{h_2})$ between the two doped holes) in CGS units in our numerical computations:
\begin{equation}\label{eq:potentials_and_fields}
	\begin{gathered}
		V_0(\va{R}_{h_1}, \va{R}_{h_2}) = \frac{|e|}{|\va{R}_{h_2} -\va{R}_{h_l}|}(1-\delta_{\va{R}_{h_2}, \va{R}_{h_l}}), \\
		V_{h,i} = |e| \sum_{l=1}^{2} \frac{1}{|\pos - \va{R}_{h_l}|} (1-\delta_{\pos, \va{R}_{h_l}}), \\
		\va{E}_{h,i} =  |e| \sum_{l=1}^{2} \frac{\pos - \va{R}_{h_l}}{|\pos - \va{R}_{h_l}|^3}(1-\delta_{\pos, \va{R}_{h_l}}), \\
		V_{p,i} = \sumions \frac{\pion \cdot (\pos - \posprime)}{|\pos - \posprime|^3}, \\
		\va{E}_{p,i} =  \sumions \frac{ 3[\pionprime \cdot (\pos - \posprime)]  (\pos - \posprime)}{|\pos - \posprime|^5} - \frac{\pionprime}{|\pos - \posprime|^3}, \\
		V_{Q,i} =   \sumions  \Qionprime \frac{1}{|\pos - \posprime|}, \\
		\va{E}_{Q,i} =  \sumions  \Qionprime \frac{\pos - \posprime}{|\pos - \posprime|^3}.
	\end{gathered}
\end{equation}
\noindent With Equation \eqref{eq:potentials_and_fields} defined, the atomic dipoles from Equation \textcolor{red}{4} can thus be written in the following way:
\begin{align}\label{eq:O_dipoles_supp}
	\pion = \alpha_i \left( \va{E}_{h,i} + \va{E}_{p,i} + \va{E}_{Q,i} \right).
\end{align}
\noindent Similarly, Equation \textcolor{red}{6} for the onsite energy can be expanded:
\begin{align}\label{eq:on-site_energy_supp}
	\epsilon_{i} = |e|\left[ V_{h,i} + V_{Q,i} + V_{p,i} \right] + U_{i},
\end{align}
\noindent where $U_{i}$ is the mean field level same-site contribution to the cost of adding a hole at $\pos$, which is elaborated upon below. 

The diagonalization of each $H_i$ Hamiltonian representing ``CuO$_4$" cluster allows us to determine the effective charge distribution of the hole each Cu ion shares with its four O neighbors in the undoped CuO$_2$ plane state. Indeed, we can extract this information from the ground states $\Phi_{i} = \mqty[\Phi^{(0)}_{i} & \Phi^{(\kappa_y)}_{i} & \Phi^{(-\kappa_x)}_{i} & \Phi^{(-\kappa_y)}_{i} & \Phi^{(\kappa_x)}_{i}]$ by considering the fact that each O ion is bonded to two Cu ions:
\begin{equation}\label{eq:charges}
	\mqty[ Q_{i} \\[0.7em] Q_{i+\kappa_y} \\[0.7em] Q_{i-\kappa_x} \\[0.7em] Q_{i-\kappa_y} \\[0.7em] Q_{i+\kappa_x} ] = |e|  \mqty[ |\Phi^{(0)}_{i}|^2 \\[0.5em] |\Phi^{(\kappa_y)}_{i}|^2 + |\Phi^{(-\kappa_y)}_{i+a\vu{y}}|^2  \\[0.5em] |\Phi^{(-\kappa_x)}_{i}|^2 + |\Phi^{(\kappa_x)}_{i-a\vu{x}}|^2 \\[0.5em] |\Phi^{(-\kappa_y)}_{i}|^2 + |\Phi^{(\kappa_y)}_{i-a\vu{y}}|^2 \\[0.5em] |\Phi^{(\kappa_x)}_{i}|^2 + |\Phi^{(-\kappa_x)}_{i+a\vu{x}}|^2 ].
\end{equation}
Finally, Equation (\textcolor{red}{8}) for the polarization energy of a given hole configuration can be written as:
\begin{align}\label{eq:energy_configuration_supp}
	E^{h}_{pol}(\va{R}_{h_1}, \va{R}_{h_2}) = \sum_{i} \biggl[ \frac{\pion}{2} \cdot \bigl( \va{E}_{Q,i} + \va{E}_{h,i} \bigr) - \Qion \left(\frac{V_{Q,i}}{2} + V_{h,i} \right) - \Omega_{i} \biggr].
\end{align}
\noindent where $\Omega_{i}$ is the interaction between charges on the same ionic site (see below).

\subsection{Same-Site Energy Definition}

The same-site energy cost that enters in Equation \eqref{eq:on-site_energy_supp} is defined as such:

\begin{align}\label{eq:U_delta}
	U_{i} = \frac{Q_{i}}{|e|} \left[ I_{i}(n_{h,i} + \gamma_{i} + 1) - I_{i}(n_{h,i} + \gamma_{i}) \right] + I_{i}(n_{h,i} + \gamma_{i}),
\end{align}

\begin{align}\label{eq:ionization_affinity}
	I_i(n) = 
	\begin{cases}
		E_{i,I}(n+1) & \text{, } n \geq 0 \\
		-E_{i,A}(|n|) & \text{, } n < 0
	\end{cases},
\end{align}

\noindent where $E_{i,I}(n)$ and $E_{i,A}(n)$ are respectively the standard n$^{th}$ atomic ionization energy and electron affinity of the element at position $i$, $\gamma_i$ is the oxidation number of the ion at $i$ ($\gamma_i$ = 1 for Cu and $\gamma_i$ = -2 for O) and $n_{h,i} = \sum^{2}_{n=1}\delta_{i, \va{R}_{h_n}}$ is the number of doped holes at site $i$. Physically, this means that the cost of adding a hole to an ion with overall charge $\gamma_i + \Qion$ will be an appropriate fraction of the relevant ionization potential or electron affinity, added to one minus that fraction of the next ionization potential or electron affinity \cite{macridin_charge_density}. For example, the cost of adding a hole to $\text{O}^{(2-) + 0.25} = \text{O}^{1.75-}$ is the sum of the costs of the steps $\text{O}^{1.75-} \rightarrow \text{O}^{1-}$ and $\text{O}^{1-} \rightarrow \text{O}^{0.75-}$, i.e. $-0.75E_{O,A}(2) - 0.25E_{O,A}(1)$. The literature energy values that end up being used in this model's numerical calculations are $E_{O,I}(1) = 13.62 \text{ eV}$ \cite{O_ionization_potential}, $E_{O,I}(2) = 35.12 \text{ eV}$ \cite{O_ionization_potential}, $E^O_A(1) = -1.46 \text{ eV}$ \cite{O_electron_affinity}, $E_{O,A}(2) = 7.71 \text{ eV}$ \cite{O_second_electron_affinity}, $E_{Cu,I}(1) = 7.72 \text{ eV}$ \cite{Cu_ionization_potential}, $E_{Cu,I}(2) = 20.29 \text{ eV}$ \cite{Cu_second_ionization_energy}, $E_{Cu,I}(3) = 36.84 \text{ eV}$ \cite{Cu_ionization_potential}.

On the other hand, the same-site term that enters in the polarization energy as featured in Equation \eqref{eq:energy_configuration_supp} is slightly different because there can be two doped holes on the same site:
\begin{align}\label{eq:Omega_energy}
	\Omega_{i} = \sum^{n_{h,i}}_{n=0} I_i(n + \gamma_i) \left[ \biggl( \frac{Q_{i}}{|e|} - 1 \biggr)\delta_{n,n_{h,i}} + 1   \right].
\end{align}
\noindent From their definitions, both $U_{i}$ and $\Omega_{i}$ are physically related to the Hubbard $U_{pp}$ and $U_{dd}$ terms describing the energy cost of having two holes (or electrons depending on the material and the context) occupying the same atomic orbital. However, they are distinct since they both represent slightly different processes; $U_{i}$ refers to the energy cost of adding one hole to an atom due to the partial charges already present and $\Omega_{i}$ is part of the polarization energy (not the global electronic energy of the system).

With all of these variables defined, the roots of a system of equations are determined numerically using the fsolve function as part of the Python scientific package SciPy \cite{SciPy}. These are then entered into Equation (\textcolor{red}{7}) of the main text to calculate the polarization energy of a given hole configuration.

\subsection{Base Charge Transfer Energy Calibration}

\begin{figure}[h]
	\includegraphics[width=\columnwidth]{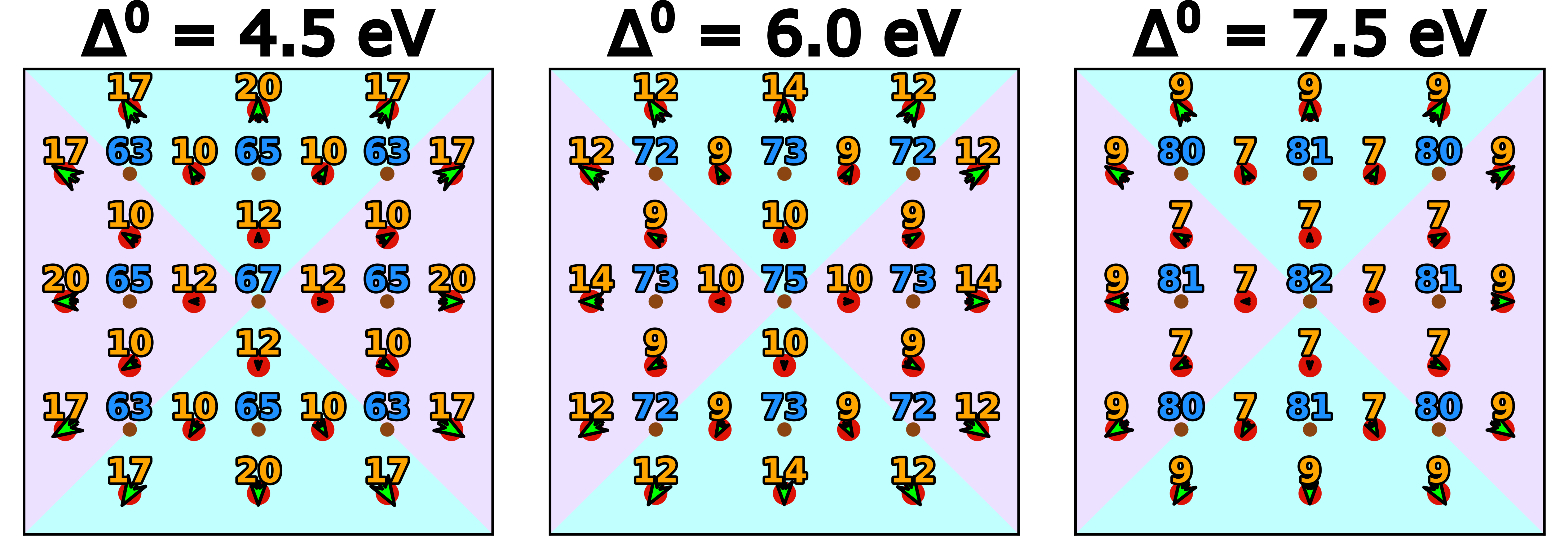}
	\caption{Ionic valence charges in an undoped CuO$_2$ plane for different values of $\Delta^0$. This figure follows the same legend as in Figure \textcolor{red}{2} in the main text. Small O$^{2-}$ atomic dipoles are present as a consequence of the finite nature of the clusters.}\label{fig:delta_choice_covalency}
\end{figure}

A disambiguation of what ``charge transfer energy'' means in this work in contrast to other models is necessary. A common value chosen to reproduce DFT band structures is 3.6 eV, but this would not be applicable to our model because in theory this number should already takes screening effects into account in order to obtain the ``real'' energy dispersion, as well as due to being used in Hamiltonians that do not explicitly consider all Coulomb interactions \cite{cuprate_review_params, cuprate_dft_params, zr_singlet}. Consequently, such models attribute the same energy cost to adding a hole on any O$^{2-}$ site (except same-site and sometimes nearest-neighbor interactions) despite the effective potential at distinct ionic sites potentially being different due to other added particles and their polarizing influence. This approximation is intrinsically related to the Clausius-Mossotti local field effect-smoothing approach that our model is going beyond. To calculate screening, we first need to determine the value of a base, undoped charge transfer energy $\Delta^0$. We calibrate the value of $\Delta^0$ such that our model leads to a proper and realistic Cu-O covalency for each hole per Cu added to the vacuum in the undoped CuO$_2$ plane. Aiming for a Cu hole density of 75\% as a compromise between the different calculations and NMR results cited in the main text, we reach as a baseline value $\Delta^0$ = 6.0 eV as shown in Figure \ref{fig:delta_choice_covalency}, specifically looking at the central Cu since edge effects are eliminated and it is influenced by other charges isotropically. Furthermore, the screened charge transfer energy $\Delta_{i, \kappa}$ we obtain for the undoped system is 1.62 eV, which is in the 1.5-2.0 eV range in which experimentally measured cuprate charge transfer energies fall \cite{eskes_delta, charge_transfer_madelung_vacuum, cuprate_experimental_delta_1.5}.

\subsection{Estimation of Reciprocal Space Approach Cost}
Another central takeaway of this work is that for calculations such as these where short-range phenomena are important to properly capture, a real space approach can offer significant computational advantages \cite{octopus_local_field}. To capture the prized second nearest-neighbor ($d = 3.80 \text{ \angstrom}$) repulsion minimum in a reciprocal space calculation, a large supercell needs to be defined depending on the desired hole doping; at the optimal 0.15 holes per Cu doping for superconductivity, we need $a = 52.20 \text{ \angstrom}$ to have 2 holes per cell \cite{cuprate_phase_diagram}. 
\begin{align}\label{eq:G-vectors}
	\sqrt{(n_x + n'_x)^2 + (n_y + n'_y)^2  + (n_z + n'_z)^2} \leq \frac{a}{d}.
\end{align}
\noindent Using Equation \eqref{eq:G-vectors} in two dimensions and only considering vectors that respect $|\va{G}_{n_x, n_y}| \leq \sqrt{2}\frac{4\pi}{d}$ leads to 3249 mandatory $\va{G}$ vectors. Similar computations in 3D materials are significantly more expensive in reciprocal space. If one is exclusively interested in what happens in the neighborhood of $\Rh = d$, including non-diagonal elements only for $\va{G}$ vectors in a small spherical shell around radius $|\va{G}_{n_x, n_y}| \approx \frac{2\pi}{d}$ while keeping the other parts of the matrix diagonal leads to size reductions. Regardless, the number of non-diagonal contributions is substantially greater for small $d$ since there are more combinations $|\va{G}_{n_x, n_y} + \va{G}_{n'_x, n'_y}| \approx \frac{2\pi}{d}$, reducing the potential savings and improving the appeal of a real space approach.


\begin{thebibliography}{68}%
	\makeatletter
	\providecommand \@ifxundefined [1]{%
		\@ifx{#1\undefined}
	}%
	\providecommand \@ifnum [1]{%
		\ifnum #1\expandafter \@firstoftwo
		\else \expandafter \@secondoftwo
		\fi
	}%
	\providecommand \@ifx [1]{%
		\ifx #1\expandafter \@firstoftwo
		\else \expandafter \@secondoftwo
		\fi
	}%
	\providecommand \natexlab [1]{#1}%
	\providecommand \enquote  [1]{``#1''}%
	\providecommand \bibnamefont  [1]{#1}%
	\providecommand \bibfnamefont [1]{#1}%
	\providecommand \citenamefont [1]{#1}%
	\providecommand \href@noop [0]{\@secondoftwo}%
	\providecommand \href [0]{\begingroup \@sanitize@url \@href}%
	\providecommand \@href[1]{\@@startlink{#1}\@@href}%
	\providecommand \@@href[1]{\endgroup#1\@@endlink}%
	\providecommand \@sanitize@url [0]{\catcode `\\12\catcode `\$12\catcode
		`\&12\catcode `\#12\catcode `\^12\catcode `\_12\catcode `\%12\relax}%
	\providecommand \@@startlink[1]{}%
	\providecommand \@@endlink[0]{}%
	\providecommand \url  [0]{\begingroup\@sanitize@url \@url }%
	\providecommand \@url [1]{\endgroup\@href {#1}{\urlprefix }}%
	\providecommand \urlprefix  [0]{URL }%
	\providecommand \Eprint [0]{\href }%
	\providecommand \doibase [0]{https://doi.org/}%
	\providecommand \selectlanguage [0]{\@gobble}%
	\providecommand \bibinfo  [0]{\@secondoftwo}%
	\providecommand \bibfield  [0]{\@secondoftwo}%
	\providecommand \translation [1]{[#1]}%
	\providecommand \BibitemOpen [0]{}%
	\providecommand \bibitemStop [0]{}%
	\providecommand \bibitemNoStop [0]{.\EOS\space}%
	\providecommand \EOS [0]{\spacefactor3000\relax}%
	\providecommand \BibitemShut  [1]{\csname bibitem#1\endcsname}%
	\let\auto@bib@innerbib\@empty
	\bibitem [{\citenamefont {Fujimori}\ \emph {et~al.}(1987)\citenamefont
		{Fujimori}, \citenamefont {Takayama-Muromachi}, \citenamefont {Uchida},\ and\
		\citenamefont {Okai}}]{cuprate_XPS_Mott}%
	\BibitemOpen
	\bibfield  {author} {\bibinfo {author} {\bibfnamefont {A.}~\bibnamefont
			{Fujimori}}, \bibinfo {author} {\bibfnamefont {E.}~\bibnamefont
			{Takayama-Muromachi}}, \bibinfo {author} {\bibfnamefont {Y.}~\bibnamefont
			{Uchida}},\ and\ \bibinfo {author} {\bibfnamefont {B.}~\bibnamefont {Okai}},\
	}\bibfield  {title} {\emph {\bibinfo {title} {Spectroscopic evidence for
				strongly correlated electronic states in {La}-{Sr}-{Cu} and {Y}-{Ba}-{Cu}
				oxides}},\ }\href {https://doi.org/10.1103/PhysRevB.35.8814} {\bibfield
		{journal} {\bibinfo  {journal} {Physical Review B}\ }\textbf {\bibinfo
			{volume} {35}},\ \bibinfo {pages} {8814--8817} (\bibinfo {year}
		{1987})}\BibitemShut {NoStop}%
	\bibitem [{\citenamefont {Yang}\ \emph {et~al.}(2023)\citenamefont {Yang},
		\citenamefont {Gao}, \citenamefont {Cao}, \citenamefont {Xu}, \citenamefont
		{Liang}, \citenamefont {Xu}, \citenamefont {Chen}, \citenamefont {Liu},
		\citenamefont {Huang}, \citenamefont {Xu}, \citenamefont {Wang},
		\citenamefont {Cui}, \citenamefont {Wang}, \citenamefont {Yang},
		\citenamefont {Luo}, \citenamefont {Sun}, \citenamefont {Yang}, \citenamefont
		{Liu},\ and\ \citenamefont {Chen}}]{4f_Mott_arpes}%
	\BibitemOpen
	\bibfield  {author} {\bibinfo {author} {\bibfnamefont {H.}~\bibnamefont
			{Yang}}, \bibinfo {author} {\bibfnamefont {J.}~\bibnamefont {Gao}}, \bibinfo
		{author} {\bibfnamefont {Y.}~\bibnamefont {Cao}}, \bibinfo {author}
		{\bibfnamefont {Y.}~\bibnamefont {Xu}}, \bibinfo {author} {\bibfnamefont
			{A.}~\bibnamefont {Liang}}, \bibinfo {author} {\bibfnamefont
			{X.}~\bibnamefont {Xu}}, \bibinfo {author} {\bibfnamefont {Y.}~\bibnamefont
			{Chen}}, \bibinfo {author} {\bibfnamefont {S.}~\bibnamefont {Liu}}, \bibinfo
		{author} {\bibfnamefont {K.}~\bibnamefont {Huang}}, \bibinfo {author}
		{\bibfnamefont {L.}~\bibnamefont {Xu}}, \bibinfo {author} {\bibfnamefont
			{C.}~\bibnamefont {Wang}}, \bibinfo {author} {\bibfnamefont {S.}~\bibnamefont
			{Cui}}, \bibinfo {author} {\bibfnamefont {M.}~\bibnamefont {Wang}}, \bibinfo
		{author} {\bibfnamefont {L.}~\bibnamefont {Yang}}, \bibinfo {author}
		{\bibfnamefont {X.}~\bibnamefont {Luo}}, \bibinfo {author} {\bibfnamefont
			{Y.}~\bibnamefont {Sun}}, \bibinfo {author} {\bibfnamefont {Y.-f.}\
			\bibnamefont {Yang}}, \bibinfo {author} {\bibfnamefont {Z.}~\bibnamefont
			{Liu}},\ and\ \bibinfo {author} {\bibfnamefont {Y.}~\bibnamefont {Chen}},\
	}\bibfield  {title} {\emph {\bibinfo {title} {Observation of {Mott}
				instability at the valence transition of f-electron system}},\ }\href
	{https://doi.org/10.1093/nsr/nwad035} {\bibfield  {journal} {\bibinfo
			{journal} {National Science Review}\ }\textbf {\bibinfo {volume} {10}},\
		\bibinfo {pages} {nwad035} (\bibinfo {year} {2023})}\BibitemShut {NoStop}%
	\bibitem [{\citenamefont {Kramida}\ \emph {et~al.}(2017)\citenamefont
		{Kramida}, \citenamefont {Nave},\ and\ \citenamefont
		{Reader}}]{Cu_second_ionization_energy}%
	\BibitemOpen
	\bibfield  {author} {\bibinfo {author} {\bibfnamefont {A.}~\bibnamefont
			{Kramida}}, \bibinfo {author} {\bibfnamefont {G.}~\bibnamefont {Nave}},\ and\
		\bibinfo {author} {\bibfnamefont {J.}~\bibnamefont {Reader}},\ }\bibfield
	{title} {\emph {\bibinfo {title} {The {Cu} {II} {Spectrum}}},\ }\href
	{https://doi.org/10.3390/atoms5010009} {\bibfield  {journal} {\bibinfo
			{journal} {Atoms}\ }\textbf {\bibinfo {volume} {5}},\ \bibinfo {pages} {9}
		(\bibinfo {year} {2017})}\BibitemShut {NoStop}%
	\bibitem [{\citenamefont {Sugar}\ and\ \citenamefont
		{Musgrove}(1990)}]{Cu_ionization_potential}%
	\BibitemOpen
	\bibfield  {author} {\bibinfo {author} {\bibfnamefont {J.}~\bibnamefont
			{Sugar}}\ and\ \bibinfo {author} {\bibfnamefont {A.}~\bibnamefont
			{Musgrove}},\ }\bibfield  {title} {\emph {\bibinfo {title} {Energy {Levels}
				of {Copper}, {Cu} {I} through {Cu} {XXIX}}},\ }\href
	{https://doi.org/10.1063/1.555855} {\bibfield  {journal} {\bibinfo  {journal}
			{Journal of Physical and Chemical Reference Data}\ }\textbf {\bibinfo
			{volume} {19}},\ \bibinfo {pages} {527--616} (\bibinfo {year}
		{1990})}\BibitemShut {NoStop}%
	\bibitem [{\citenamefont {de~Boer}\ \emph {et~al.}(1984)\citenamefont
		{de~Boer}, \citenamefont {Haas},\ and\ \citenamefont
		{Sawatzky}}]{anion_screening_sawatzky}%
	\BibitemOpen
	\bibfield  {author} {\bibinfo {author} {\bibfnamefont {D.~K.~G.}\
			\bibnamefont {de~Boer}}, \bibinfo {author} {\bibfnamefont {C.}~\bibnamefont
			{Haas}},\ and\ \bibinfo {author} {\bibfnamefont {G.~A.}\ \bibnamefont
			{Sawatzky}},\ }\bibfield  {title} {\emph {\bibinfo {title} {Exciton
				satellites in photoelectron spectra}},\ }\href
	{https://doi.org/10.1103/PhysRevB.29.4401} {\bibfield  {journal} {\bibinfo
			{journal} {Physical Review B}\ }\textbf {\bibinfo {volume} {29}},\ \bibinfo
		{pages} {4401--4419} (\bibinfo {year} {1984})},\ \bibinfo {note} {publisher:
		American Physical Society}\BibitemShut {NoStop}%
	\bibitem [{\citenamefont {Vaugier}\ \emph {et~al.}(2012)\citenamefont
		{Vaugier}, \citenamefont {Jiang},\ and\ \citenamefont
		{Biermann}}]{srvo3_hubbard_u}%
	\BibitemOpen
	\bibfield  {author} {\bibinfo {author} {\bibfnamefont {L.}~\bibnamefont
			{Vaugier}}, \bibinfo {author} {\bibfnamefont {H.}~\bibnamefont {Jiang}},\
		and\ \bibinfo {author} {\bibfnamefont {S.}~\bibnamefont {Biermann}},\
	}\bibfield  {title} {\emph {\bibinfo {title} {Hubbard {U} and {Hund} exchange
				{J} in transition metal oxides: {Screening} versus localization trends from
				constrained random phase approximation}},\ }\href
	{https://doi.org/10.1103/PhysRevB.86.165105} {\bibfield  {journal} {\bibinfo
			{journal} {Physical Review B}\ }\textbf {\bibinfo {volume} {86}},\ \bibinfo
		{pages} {165105} (\bibinfo {year} {2012})},\ \bibinfo {note} {publisher:
		American Physical Society}\BibitemShut {NoStop}%
	\bibitem [{\citenamefont {van~der Marel}\ \emph {et~al.}(1988)\citenamefont
		{van~der Marel}, \citenamefont {van Elp}, \citenamefont {Sawatzky},\ and\
		\citenamefont {Heitmann}}]{cuprate_sawatzky_xps_udd}%
	\BibitemOpen
	\bibfield  {author} {\bibinfo {author} {\bibfnamefont {D.}~\bibnamefont
			{van~der Marel}}, \bibinfo {author} {\bibfnamefont {J.}~\bibnamefont {van
				Elp}}, \bibinfo {author} {\bibfnamefont {G.~A.}\ \bibnamefont {Sawatzky}},\
		and\ \bibinfo {author} {\bibfnamefont {D.}~\bibnamefont {Heitmann}},\
	}\bibfield  {title} {\emph {\bibinfo {title} {X-ray photoemission,
				bremsstrahlung isochromat, {Auger}-electron, and optical spectroscopy studies
				of {Y}-{Ba}-{Cu}-{O} thin films}},\ }\href
	{https://doi.org/10.1103/PhysRevB.37.5136} {\bibfield  {journal} {\bibinfo
			{journal} {Physical Review B}\ }\textbf {\bibinfo {volume} {37}},\ \bibinfo
		{pages} {5136--5141} (\bibinfo {year} {1988})},\ \bibinfo {note} {publisher:
		American Physical Society}\BibitemShut {NoStop}%
	\bibitem [{\citenamefont {Clausius}(1879)}]{Clausius_original}%
	\BibitemOpen
	\bibfield  {author} {\bibinfo {author} {\bibfnamefont {R.}~\bibnamefont
			{Clausius}},\ }\href {https://doi.org/10.1007/978-3-663-20232-5} {\emph
		{\bibinfo {title} {Die {Mechanische} {Behandlung} der {Electricität}}}}\
	(\bibinfo  {publisher} {Vieweg+Teubner Verlag},\ \bibinfo {address}
	{Wiesbaden},\ \bibinfo {year} {1879})\BibitemShut {NoStop}%
	\bibitem [{\citenamefont {Mossotti}(1850)}]{mossotti_original}%
	\BibitemOpen
	\bibfield  {author} {\bibinfo {author} {\bibfnamefont {O.~F.}\ \bibnamefont
			{Mossotti}},\ }in\ \href
	{https://www.biodiversitylibrary.org/item/34306#page/1/mode/1up} {\emph
		{\bibinfo {booktitle} {Memorie di {Matematica} e di {Fisica} della {Società}
				{Italiana} delle {Scienze} {Residente} in {Modena}}}},\ Vol.~\bibinfo
	{volume} {24}\ (\bibinfo  {publisher} {Società Italiana Delle Scienze},\
	\bibinfo {year} {1850})\ pp.\ \bibinfo {pages} {49--74}\BibitemShut {NoStop}%
	\bibitem [{\citenamefont {Sawatzky}\ \emph {et~al.}(2009)\citenamefont
		{Sawatzky}, \citenamefont {Elfimov}, \citenamefont {Van Den~Brink},\ and\
		\citenamefont {Zaanen}}]{FeAs_Sawatzky}%
	\BibitemOpen
	\bibfield  {author} {\bibinfo {author} {\bibfnamefont {G.~A.}\ \bibnamefont
			{Sawatzky}}, \bibinfo {author} {\bibfnamefont {I.~S.}\ \bibnamefont
			{Elfimov}}, \bibinfo {author} {\bibfnamefont {J.}~\bibnamefont {Van
				Den~Brink}},\ and\ \bibinfo {author} {\bibfnamefont {J.}~\bibnamefont
			{Zaanen}},\ }\bibfield  {title} {\emph {\bibinfo {title} {Heavy-anion
				solvation of polarity fluctuations in pnictides}},\ }\bibfield  {journal}
	{\bibinfo  {journal} {Epl}\ }\textbf {\bibinfo {volume} {86}},\ \href
	{https://doi.org/10.1209/0295-5075/86/17006} {10.1209/0295-5075/86/17006}
	(\bibinfo {year} {2009}),\ \bibinfo {note} {arXiv: 0808.1390}\BibitemShut
	{NoStop}%
	\bibitem [{\citenamefont {Brocks}\ \emph {et~al.}(2004)\citenamefont {Brocks},
		\citenamefont {van~den Brink},\ and\ \citenamefont
		{Morpurgo}}]{jeroen_screening_organic}%
	\BibitemOpen
	\bibfield  {author} {\bibinfo {author} {\bibfnamefont {G.}~\bibnamefont
			{Brocks}}, \bibinfo {author} {\bibfnamefont {J.}~\bibnamefont {van~den
				Brink}},\ and\ \bibinfo {author} {\bibfnamefont {A.~F.}\ \bibnamefont
			{Morpurgo}},\ }\bibfield  {title} {\emph {\bibinfo {title} {Electronic
				{Correlations} in {Oligo}-acene and -{Thiopene} {Organic} {Molecular}
				{Crystals}}},\ }\href {https://doi.org/10.1103/PhysRevLett.93.146405}
	{\bibfield  {journal} {\bibinfo  {journal} {Physical Review Letters}\
		}\textbf {\bibinfo {volume} {93}},\ \bibinfo {pages} {146405} (\bibinfo
		{year} {2004})}\BibitemShut {NoStop}%
	\bibitem [{\citenamefont {Caprara}\ \emph {et~al.}(2017)\citenamefont
		{Caprara}, \citenamefont {Grilli}, \citenamefont {Di~Castro},\ and\
		\citenamefont {Seibold}}]{Caprara_cuprate_cdw_anisotropy}%
	\BibitemOpen
	\bibfield  {author} {\bibinfo {author} {\bibfnamefont {S.}~\bibnamefont
			{Caprara}}, \bibinfo {author} {\bibfnamefont {M.}~\bibnamefont {Grilli}},
		\bibinfo {author} {\bibfnamefont {C.}~\bibnamefont {Di~Castro}},\ and\
		\bibinfo {author} {\bibfnamefont {G.}~\bibnamefont {Seibold}},\ }\bibfield
	{title} {\emph {\bibinfo {title} {Pseudogap and ({An})isotropic {Scattering}
				in the {Fluctuating} {Charge}-{Density} {Wave} {Phase} of {Cuprates}}},\
	}\href {https://doi.org/10.1007/s10948-016-3775-9} {\bibfield  {journal}
		{\bibinfo  {journal} {Journal of Superconductivity and Novel Magnetism}\
		}\textbf {\bibinfo {volume} {30}},\ \bibinfo {pages} {25--30} (\bibinfo
		{year} {2017})}\BibitemShut {NoStop}%
	\bibitem [{\citenamefont {Friedel}(1958)}]{Friedel_original}%
	\BibitemOpen
	\bibfield  {author} {\bibinfo {author} {\bibfnamefont {J.}~\bibnamefont
			{Friedel}},\ }\bibfield  {title} {\emph {\bibinfo {title} {Metallic
				alloys}},\ }\href {https://doi.org/10.1007/BF02751483} {\bibfield  {journal}
		{\bibinfo  {journal} {Il Nuovo Cimento (1955-1965)}\ }\textbf {\bibinfo
			{volume} {7}},\ \bibinfo {pages} {287--311} (\bibinfo {year}
		{1958})}\BibitemShut {NoStop}%
	\bibitem [{\citenamefont {Hammel}\ \emph {et~al.}(1989)\citenamefont {Hammel},
		\citenamefont {Takigawa}, \citenamefont {Heffner}, \citenamefont {Fisk},\
		and\ \citenamefont {Ott}}]{nmr_cuprates}%
	\BibitemOpen
	\bibfield  {author} {\bibinfo {author} {\bibfnamefont {P.}~\bibnamefont
			{Hammel}}, \bibinfo {author} {\bibfnamefont {M.}~\bibnamefont {Takigawa}},
		\bibinfo {author} {\bibfnamefont {R.}~\bibnamefont {Heffner}}, \bibinfo
		{author} {\bibfnamefont {Z.}~\bibnamefont {Fisk}},\ and\ \bibinfo {author}
		{\bibfnamefont {K.}~\bibnamefont {Ott}},\ }\bibfield  {title} {\emph
		{\bibinfo {title} {Nuclear relaxation rates at copper and oxygen sites in
				{YBa}$_{\textrm{2}}${Cu}$_{\textrm{3}}${O}$_{\textrm{7}}$}},\ }\href
	{https://doi.org/10.1016/0921-4534(89)90975-1} {\bibfield  {journal}
		{\bibinfo  {journal} {Physica C: Superconductivity and its Applications}\
		}\textbf {\bibinfo {volume} {162-164}},\ \bibinfo {pages} {177--178}
		(\bibinfo {year} {1989})}\BibitemShut {NoStop}%
	\bibitem [{\citenamefont {Macridin}\ \emph {et~al.}(2005)\citenamefont
		{Macridin}, \citenamefont {Jarrell}, \citenamefont {Maier},\ and\
		\citenamefont {Sawatzky}}]{macridin_charge_density}%
	\BibitemOpen
	\bibfield  {author} {\bibinfo {author} {\bibfnamefont {A.}~\bibnamefont
			{Macridin}}, \bibinfo {author} {\bibfnamefont {M.}~\bibnamefont {Jarrell}},
		\bibinfo {author} {\bibfnamefont {T.}~\bibnamefont {Maier}},\ and\ \bibinfo
		{author} {\bibfnamefont {G.~A.}\ \bibnamefont {Sawatzky}},\ }\bibfield
	{title} {\emph {\bibinfo {title} {Physics of cuprates with the two-band
				{Hubbard} model: {The} validity of the one-band {Hubbard} model}},\
	}\bibfield  {journal} {\bibinfo  {journal} {Physical Review B}\ }\textbf
	{\bibinfo {volume} {71}},\ \href {https://doi.org/10.1103/PhysRevB.71.134527}
	{10.1103/PhysRevB.71.134527} (\bibinfo {year} {2005}),\ \bibinfo {note}
	{arXiv: cond-mat/0411092}\BibitemShut {NoStop}%
	\bibitem [{\citenamefont {Taeck~Park}\ \emph {et~al.}(1988)\citenamefont
		{Taeck~Park}, \citenamefont {Terakura}, \citenamefont {Oguchi}, \citenamefont
		{Yanase},\ and\ \citenamefont {Ikeda}}]{lda_cuprate_correlated}%
	\BibitemOpen
	\bibfield  {author} {\bibinfo {author} {\bibfnamefont {K.}~\bibnamefont
			{Taeck~Park}}, \bibinfo {author} {\bibfnamefont {K.}~\bibnamefont
			{Terakura}}, \bibinfo {author} {\bibfnamefont {T.}~\bibnamefont {Oguchi}},
		\bibinfo {author} {\bibfnamefont {A.}~\bibnamefont {Yanase}},\ and\ \bibinfo
		{author} {\bibfnamefont {M.}~\bibnamefont {Ikeda}},\ }\bibfield  {title}
	{\emph {\bibinfo {title} {Implications of {Band}-{Structure} {Calculations}
				for {High}-{Tc} {Related} {Oxides}}},\ }\href
	{https://doi.org/10.1143/JPSJ.57.3445} {\bibfield  {journal} {\bibinfo
			{journal} {Journal of the Physical Society of Japan}\ }\textbf {\bibinfo
			{volume} {57}},\ \bibinfo {pages} {3445--3456} (\bibinfo {year} {1988})},\
	\bibinfo {note} {publisher: The Physical Society of Japan}\BibitemShut
	{NoStop}%
	\bibitem [{\citenamefont {Pines}\ and\ \citenamefont
		{Bohm}(1952)}]{rpa_original_2}%
	\BibitemOpen
	\bibfield  {author} {\bibinfo {author} {\bibfnamefont {D.}~\bibnamefont
			{Pines}}\ and\ \bibinfo {author} {\bibfnamefont {D.}~\bibnamefont {Bohm}},\
	}\bibfield  {title} {\emph {\bibinfo {title} {A {Collective} {Description} of
				{Electron} {Interactions}: {II}. {Collective} vs {Individual} {Particle}
				{Aspects} of the {Interactions}}},\ }\href
	{https://doi.org/10.1103/PhysRev.85.338} {\bibfield  {journal} {\bibinfo
			{journal} {Physical Review}\ }\textbf {\bibinfo {volume} {85}},\ \bibinfo
		{pages} {338--353} (\bibinfo {year} {1952})},\ \bibinfo {note} {publisher:
		American Physical Society}\BibitemShut {NoStop}%
	\bibitem [{\citenamefont {Mazin}\ and\ \citenamefont
		{Cohen}(1997)}]{RPA_dft_mazin}%
	\BibitemOpen
	\bibfield  {author} {\bibinfo {author} {\bibfnamefont {I.~I.}\ \bibnamefont
			{Mazin}}\ and\ \bibinfo {author} {\bibfnamefont {R.~E.}\ \bibnamefont
			{Cohen}},\ }\bibfield  {title} {\emph {\bibinfo {title} {Notes on the static
				dielectric response function in the density functional theory}},\ }\bibfield
	{journal} {\bibinfo  {journal} {Ferroelectrics}\ }\href
	{https://doi.org/10.1080/00150199708016098} {10.1080/00150199708016098}
	(\bibinfo {year} {1997})\BibitemShut {NoStop}%
	\bibitem [{\citenamefont {Zhang}\ and\ \citenamefont
		{Rice}(1988)}]{zr_singlet}%
	\BibitemOpen
	\bibfield  {author} {\bibinfo {author} {\bibfnamefont {F.~C.}\ \bibnamefont
			{Zhang}}\ and\ \bibinfo {author} {\bibfnamefont {T.~M.}\ \bibnamefont
			{Rice}},\ }\bibfield  {title} {\emph {\bibinfo {title} {Effective
				{Hamiltonian} for the superconducting {Cu} oxides}},\ }\href
	{https://doi.org/10.1103/PhysRevB.37.3759} {\bibfield  {journal} {\bibinfo
			{journal} {Physical Review B}\ }\textbf {\bibinfo {volume} {37}},\ \bibinfo
		{pages} {3759--3761} (\bibinfo {year} {1988})}\BibitemShut {NoStop}%
	\bibitem [{\citenamefont {Berciu}\ \emph {et~al.}(2009)\citenamefont {Berciu},
		\citenamefont {Elfimov},\ and\ \citenamefont
		{Sawatzky}}]{sawatzky_pnictides}%
	\BibitemOpen
	\bibfield  {author} {\bibinfo {author} {\bibfnamefont {M.}~\bibnamefont
			{Berciu}}, \bibinfo {author} {\bibfnamefont {I.}~\bibnamefont {Elfimov}},\
		and\ \bibinfo {author} {\bibfnamefont {G.~A.}\ \bibnamefont {Sawatzky}},\
	}\bibfield  {title} {\emph {\bibinfo {title} {Electronic polarons and
				bipolarons in iron-based superconductors: {The} role of anions}},\ }\href
	{https://doi.org/10.1103/PhysRevB.79.214507} {\bibfield  {journal} {\bibinfo
			{journal} {Physical Review B}\ }\textbf {\bibinfo {volume} {79}},\ \bibinfo
		{pages} {214507} (\bibinfo {year} {2009})}\BibitemShut {NoStop}%
	\bibitem [{\citenamefont {King-Smith}\ and\ \citenamefont
		{Vanderbilt}(1993)}]{modern_theory_of_polarization_precursor}%
	\BibitemOpen
	\bibfield  {author} {\bibinfo {author} {\bibfnamefont {R.~D.}\ \bibnamefont
			{King-Smith}}\ and\ \bibinfo {author} {\bibfnamefont {D.}~\bibnamefont
			{Vanderbilt}},\ }\bibfield  {title} {\emph {\bibinfo {title} {Theory of
				polarization of crystalline solids}},\ }\href
	{https://doi.org/10.1103/PhysRevB.47.1651} {\bibfield  {journal} {\bibinfo
			{journal} {Physical Review B}\ }\textbf {\bibinfo {volume} {47}},\ \bibinfo
		{pages} {1651--1654} (\bibinfo {year} {1993})},\ \bibinfo {note} {publisher:
		American Physical Society}\BibitemShut {NoStop}%
	\bibitem [{\citenamefont
		{Resta}(1993)}]{modern_theory_of_polarization_original}%
	\BibitemOpen
	\bibfield  {author} {\bibinfo {author} {\bibfnamefont {R.}~\bibnamefont
			{Resta}},\ }\bibfield  {title} {\emph {\bibinfo {title} {Macroscopic
				{Electric} {Polarization} as a {Geometric} {Quantum} {Phase}}},\ }\href
	{https://doi.org/10.1209/0295-5075/22/2/010} {\bibfield  {journal} {\bibinfo
			{journal} {Europhysics Letters}\ }\textbf {\bibinfo {volume} {22}},\ \bibinfo
		{pages} {133} (\bibinfo {year} {1993})}\BibitemShut {NoStop}%
	\bibitem [{\citenamefont {Emery}(1987)}]{emery_cuprate}%
	\BibitemOpen
	\bibfield  {author} {\bibinfo {author} {\bibfnamefont {V.~J.}\ \bibnamefont
			{Emery}},\ }\bibfield  {title} {\emph {\bibinfo {title} {Theory of
				high-{T}$_{\textrm{c}}$ superconductivity in oxides}},\ }\href
	{https://doi.org/10.1103/PhysRevLett.58.2794} {\bibfield  {journal} {\bibinfo
			{journal} {Physical Review Letters}\ }\textbf {\bibinfo {volume} {58}},\
		\bibinfo {pages} {2794--2797} (\bibinfo {year} {1987})}\BibitemShut {NoStop}%
	\bibitem [{\citenamefont {Zaanen}\ \emph {et~al.}(1985)\citenamefont {Zaanen},
		\citenamefont {Sawatzky},\ and\ \citenamefont {Allen}}]{zsa_original_paper}%
	\BibitemOpen
	\bibfield  {author} {\bibinfo {author} {\bibfnamefont {J.}~\bibnamefont
			{Zaanen}}, \bibinfo {author} {\bibfnamefont {G.~A.}\ \bibnamefont
			{Sawatzky}},\ and\ \bibinfo {author} {\bibfnamefont {J.~W.}\ \bibnamefont
			{Allen}},\ }\bibfield  {title} {\emph {\bibinfo {title} {Band gaps and
				electronic structure of transition-metal compounds}},\ }\href
	{https://doi.org/10.1103/PhysRevLett.55.418} {\bibfield  {journal} {\bibinfo
			{journal} {Physical Review Letters}\ }\textbf {\bibinfo {volume} {55}},\
		\bibinfo {pages} {418--421} (\bibinfo {year} {1985})}\BibitemShut {NoStop}%
	\bibitem [{\citenamefont {Zaanen}\ and\ \citenamefont
		{Sawatzky}(1990)}]{sawatzky_hubbard_u}%
	\BibitemOpen
	\bibfield  {author} {\bibinfo {author} {\bibfnamefont {J.}~\bibnamefont
			{Zaanen}}\ and\ \bibinfo {author} {\bibfnamefont {G.}~\bibnamefont
			{Sawatzky}},\ }\bibfield  {title} {\emph {\bibinfo {title} {Systematics in
				band gaps and optical spectra of {3D} transition metal compounds}},\ }\href
	{https://doi.org/10.1016/0022-4596(90)90202-9} {\bibfield  {journal}
		{\bibinfo  {journal} {Journal of Solid State Chemistry}\ }\textbf {\bibinfo
			{volume} {88}},\ \bibinfo {pages} {8--27} (\bibinfo {year}
		{1990})}\BibitemShut {NoStop}%
	\bibitem [{\citenamefont {McMahan}\ \emph {et~al.}(1990)\citenamefont
		{McMahan}, \citenamefont {Annett},\ and\ \citenamefont
		{Martin}}]{cuprate_dft_params}%
	\BibitemOpen
	\bibfield  {author} {\bibinfo {author} {\bibfnamefont {A.~K.}\ \bibnamefont
			{McMahan}}, \bibinfo {author} {\bibfnamefont {J.~F.}\ \bibnamefont
			{Annett}},\ and\ \bibinfo {author} {\bibfnamefont {R.~M.}\ \bibnamefont
			{Martin}},\ }\bibfield  {title} {\emph {\bibinfo {title} {Cuprate parameters
				from numerical {Wannier} functions}},\ }\href
	{https://doi.org/10.1103/PhysRevB.42.6268} {\bibfield  {journal} {\bibinfo
			{journal} {Physical Review B}\ }\textbf {\bibinfo {volume} {42}},\ \bibinfo
		{pages} {6268--6282} (\bibinfo {year} {1990})}\BibitemShut {NoStop}%
	\bibitem[{sup()}]{supp_material_cuprate}
	\bibinfo{note}{See Supplemental Material for further detailed
		description of the polarization model, specifically the electrostatics
		equations used for the fields and potentials, the definition of the on-site
		energies used and the calibration of the charge transfer energy, which
		includes Refs. [28-35].}
	\BibitemOpen
	\bibitem [{\citenamefont {Kelly}\ and\ \citenamefont
		{Palumbo}(1973)}]{O_ionization_potential}%
	\BibitemOpen
	\bibfield  {author} {\bibinfo {author} {\bibfnamefont {R.~L.}\ \bibnamefont
			{Kelly}}\ and\ \bibinfo {author} {\bibfnamefont {L.~J.}\ \bibnamefont
			{Palumbo}},\ }\href {https://www.osti.gov/biblio/4332843} {\emph {\bibinfo
			{title} {Atomic and ionic emission lines below 2000 angstroms: hydrogen
				through krypton}}},\ \bibinfo {type} {Tech. {Rep}.}\ \bibinfo {number}
	{NRL-7599}\ (\bibinfo  {institution} {Naval Research Lab., Washington, D.C.
		(USA)},\ \bibinfo {year} {1973})\BibitemShut {NoStop}%
	\bibitem [{\citenamefont {Kristiansson}\ \emph {et~al.}(2022)\citenamefont
		{Kristiansson}, \citenamefont {Chartkunchand}, \citenamefont {Eklund},
		\citenamefont {Hole}, \citenamefont {Anderson}, \citenamefont {de~Ruette},
		\citenamefont {Kamińska}, \citenamefont {Punnakayathil}, \citenamefont
		{Navarro-Navarrete}, \citenamefont {Sigurdsson}, \citenamefont {Grumer},
		\citenamefont {Simonsson}, \citenamefont {Björkhage}, \citenamefont
		{Rosén}, \citenamefont {Reinhed}, \citenamefont {Blom}, \citenamefont
		{Källberg}, \citenamefont {Alexander}, \citenamefont {Cederquist},
		\citenamefont {Zettergren}, \citenamefont {Schmidt},\ and\ \citenamefont
		{Hanstorp}}]{O_electron_affinity}%
	\BibitemOpen
	\bibfield  {author} {\bibinfo {author} {\bibfnamefont {M.~K.}\ \bibnamefont
			{Kristiansson}}, \bibinfo {author} {\bibfnamefont {K.}~\bibnamefont
			{Chartkunchand}}, \bibinfo {author} {\bibfnamefont {G.}~\bibnamefont
			{Eklund}}, \bibinfo {author} {\bibfnamefont {O.~M.}\ \bibnamefont {Hole}},
		\bibinfo {author} {\bibfnamefont {E.~K.}\ \bibnamefont {Anderson}}, \bibinfo
		{author} {\bibfnamefont {N.}~\bibnamefont {de~Ruette}}, \bibinfo {author}
		{\bibfnamefont {M.}~\bibnamefont {Kamińska}}, \bibinfo {author}
		{\bibfnamefont {N.}~\bibnamefont {Punnakayathil}}, \bibinfo {author}
		{\bibfnamefont {J.~E.}\ \bibnamefont {Navarro-Navarrete}}, \bibinfo {author}
		{\bibfnamefont {S.}~\bibnamefont {Sigurdsson}}, \bibinfo {author}
		{\bibfnamefont {J.}~\bibnamefont {Grumer}}, \bibinfo {author} {\bibfnamefont
			{A.}~\bibnamefont {Simonsson}}, \bibinfo {author} {\bibfnamefont
			{M.}~\bibnamefont {Björkhage}}, \bibinfo {author} {\bibfnamefont
			{S.}~\bibnamefont {Rosén}}, \bibinfo {author} {\bibfnamefont
			{P.}~\bibnamefont {Reinhed}}, \bibinfo {author} {\bibfnamefont
			{M.}~\bibnamefont {Blom}}, \bibinfo {author} {\bibfnamefont {A.}~\bibnamefont
			{Källberg}}, \bibinfo {author} {\bibfnamefont {J.~D.}\ \bibnamefont
			{Alexander}}, \bibinfo {author} {\bibfnamefont {H.}~\bibnamefont
			{Cederquist}}, \bibinfo {author} {\bibfnamefont {H.}~\bibnamefont
			{Zettergren}}, \bibinfo {author} {\bibfnamefont {H.~T.}\ \bibnamefont
			{Schmidt}},\ and\ \bibinfo {author} {\bibfnamefont {D.}~\bibnamefont
			{Hanstorp}},\ }\bibfield  {title} {\emph {\bibinfo {title} {High-precision
				electron affinity of oxygen}},\ }\href
	{https://doi.org/10.1038/s41467-022-33438-y} {\bibfield  {journal} {\bibinfo
			{journal} {Nature Communications}\ }\textbf {\bibinfo {volume} {13}},\
		\bibinfo {pages} {5906} (\bibinfo {year} {2022})}\BibitemShut {NoStop}%
	\bibitem [{\citenamefont {James E.~Huheey}(1993)}]{O_second_electron_affinity}%
	\BibitemOpen
	\bibfield  {author} {\bibinfo {author} {\bibfnamefont {R.~L.~K.}\
			\bibnamefont {James E.~Huheey}, \bibfnamefont {Ellen A.~Keiter}},\ }\href
	{https://www.semanticscholar.org/paper/Inorganic-chemistry%3B-principles-of-structure-and-Huheey/c0f189ba30e082b6ddfa98dc03dbd1d17d39f1b6}
	{\emph {\bibinfo {title} {Inorganic {Chemistry} {Principles} of {Structure}
				and {Relativity}}}}\ (\bibinfo  {publisher} {HarperCollins},\ \bibinfo
	{address} {New York},\ \bibinfo {year} {1993})\BibitemShut {NoStop}%
	\bibitem [{\citenamefont {Virtanen}\ \emph {et~al.}(2020)\citenamefont
		{Virtanen}, \citenamefont {Gommers}, \citenamefont {Oliphant}, \citenamefont
		{Haberland}, \citenamefont {Reddy}, \citenamefont {Cournapeau}, \citenamefont
		{Burovski}, \citenamefont {Peterson}, \citenamefont {Weckesser},
		\citenamefont {Bright}, \citenamefont {van~der Walt}, \citenamefont {Brett},
		\citenamefont {Wilson}, \citenamefont {Millman}, \citenamefont {Mayorov},
		\citenamefont {Nelson}, \citenamefont {Jones}, \citenamefont {Kern},
		\citenamefont {Larson}, \citenamefont {Carey}, \citenamefont {Polat},
		\citenamefont {Feng}, \citenamefont {Moore}, \citenamefont {VanderPlas},
		\citenamefont {Laxalde}, \citenamefont {Perktold}, \citenamefont {Cimrman},
		\citenamefont {Henriksen}, \citenamefont {Quintero}, \citenamefont {Harris},
		\citenamefont {Archibald}, \citenamefont {Ribeiro}, \citenamefont
		{Pedregosa},\ and\ \citenamefont {van Mulbregt}}]{SciPy}%
	\BibitemOpen
	\bibfield  {author} {\bibinfo {author} {\bibfnamefont {P.}~\bibnamefont
			{Virtanen}}, \bibinfo {author} {\bibfnamefont {R.}~\bibnamefont {Gommers}},
		\bibinfo {author} {\bibfnamefont {T.~E.}\ \bibnamefont {Oliphant}}, \bibinfo
		{author} {\bibfnamefont {M.}~\bibnamefont {Haberland}}, \bibinfo {author}
		{\bibfnamefont {T.}~\bibnamefont {Reddy}}, \bibinfo {author} {\bibfnamefont
			{D.}~\bibnamefont {Cournapeau}}, \bibinfo {author} {\bibfnamefont
			{E.}~\bibnamefont {Burovski}}, \bibinfo {author} {\bibfnamefont
			{P.}~\bibnamefont {Peterson}}, \bibinfo {author} {\bibfnamefont
			{W.}~\bibnamefont {Weckesser}}, \bibinfo {author} {\bibfnamefont
			{J.}~\bibnamefont {Bright}}, \bibinfo {author} {\bibfnamefont {S.~J.}\
			\bibnamefont {van~der Walt}}, \bibinfo {author} {\bibfnamefont
			{M.}~\bibnamefont {Brett}}, \bibinfo {author} {\bibfnamefont
			{J.}~\bibnamefont {Wilson}}, \bibinfo {author} {\bibfnamefont {K.~J.}\
			\bibnamefont {Millman}}, \bibinfo {author} {\bibfnamefont {N.}~\bibnamefont
			{Mayorov}}, \bibinfo {author} {\bibfnamefont {A.~R.~J.}\ \bibnamefont
			{Nelson}}, \bibinfo {author} {\bibfnamefont {E.}~\bibnamefont {Jones}},
		\bibinfo {author} {\bibfnamefont {R.}~\bibnamefont {Kern}}, \bibinfo {author}
		{\bibfnamefont {E.}~\bibnamefont {Larson}}, \bibinfo {author} {\bibfnamefont
			{C.~J.}\ \bibnamefont {Carey}}, \bibinfo {author} {\bibfnamefont
			{I.}~\bibnamefont {Polat}}, \bibinfo {author} {\bibfnamefont
			{Y.}~\bibnamefont {Feng}}, \bibinfo {author} {\bibfnamefont {E.~W.}\
			\bibnamefont {Moore}}, \bibinfo {author} {\bibfnamefont {J.}~\bibnamefont
			{VanderPlas}}, \bibinfo {author} {\bibfnamefont {D.}~\bibnamefont {Laxalde}},
		\bibinfo {author} {\bibfnamefont {J.}~\bibnamefont {Perktold}}, \bibinfo
		{author} {\bibfnamefont {R.}~\bibnamefont {Cimrman}}, \bibinfo {author}
		{\bibfnamefont {I.}~\bibnamefont {Henriksen}}, \bibinfo {author}
		{\bibfnamefont {E.~A.}\ \bibnamefont {Quintero}}, \bibinfo {author}
		{\bibfnamefont {C.~R.}\ \bibnamefont {Harris}}, \bibinfo {author}
		{\bibfnamefont {A.~M.}\ \bibnamefont {Archibald}}, \bibinfo {author}
		{\bibfnamefont {A.~H.}\ \bibnamefont {Ribeiro}}, \bibinfo {author}
		{\bibfnamefont {F.}~\bibnamefont {Pedregosa}},\ and\ \bibinfo {author}
		{\bibfnamefont {P.}~\bibnamefont {van Mulbregt}},\ }\bibfield  {title} {\emph
		{\bibinfo {title} {{SciPy} 1.0: fundamental algorithms for scientific
				computing in {Python}}},\ }\href {https://doi.org/10.1038/s41592-019-0686-2}
	{\bibfield  {journal} {\bibinfo  {journal} {Nature Methods}\ }\textbf
		{\bibinfo {volume} {17}},\ \bibinfo {pages} {261--272} (\bibinfo {year}
		{2020})},\ \bibinfo {note} {publisher: Nature Publishing Group}\BibitemShut
	{NoStop}%
	\bibitem [{\citenamefont {Tsukada}\ \emph {et~al.}(2006)\citenamefont
		{Tsukada}, \citenamefont {Shibata}, \citenamefont {Noda}, \citenamefont
		{Yamamoto},\ and\ \citenamefont {Naito}}]{charge_transfer_madelung_vacuum}%
	\BibitemOpen
	\bibfield  {author} {\bibinfo {author} {\bibfnamefont {A.}~\bibnamefont
			{Tsukada}}, \bibinfo {author} {\bibfnamefont {H.}~\bibnamefont {Shibata}},
		\bibinfo {author} {\bibfnamefont {M.}~\bibnamefont {Noda}}, \bibinfo {author}
		{\bibfnamefont {H.}~\bibnamefont {Yamamoto}},\ and\ \bibinfo {author}
		{\bibfnamefont {M.}~\bibnamefont {Naito}},\ }\bibfield  {title} {\emph
		{\bibinfo {title} {Charge transfer gap for
				{T}'-{RE}$_{\textrm{2}}${CuO}$_{\textrm{4}}$ and
				{T}-{La}$_{\textrm{2}}${CuO}$_{\textrm{4}}$ as estimated from {Madelung}
				potential calculations}},\ }\href
	{https://doi.org/10.1016/j.physc.2006.03.087} {\bibfield  {journal} {\bibinfo
			{journal} {Physica C: Superconductivity and its Applications}\ }\textbf
		{\bibinfo {volume} {445-448}},\ \bibinfo {pages} {94--96} (\bibinfo {year}
		{2006})}\BibitemShut {NoStop}%
	\bibitem [{\citenamefont {Stechel}\ and\ \citenamefont
		{Jennison}(1988)}]{cuprate_experimental_delta_1.5}%
	\BibitemOpen
	\bibfield  {author} {\bibinfo {author} {\bibfnamefont {E.~B.}\ \bibnamefont
			{Stechel}}\ and\ \bibinfo {author} {\bibfnamefont {D.~R.}\ \bibnamefont
			{Jennison}},\ }\bibfield  {title} {\emph {\bibinfo {title} {Electronic
				structure of {CuO}$_{\textrm{2}}$ sheets and spin-driven high-{Tc}
				superconductivity}},\ }\href {https://doi.org/10.1103/PhysRevB.38.4632}
	{\bibfield  {journal} {\bibinfo  {journal} {Physical Review B}\ }\textbf
		{\bibinfo {volume} {38}},\ \bibinfo {pages} {4632--4659} (\bibinfo {year}
		{1988})}\BibitemShut {NoStop}%
	\bibitem [{\citenamefont {Castro}\ \emph {et~al.}(2006)\citenamefont {Castro},
		\citenamefont {Appel}, \citenamefont {Oliveira}, \citenamefont {Rozzi},
		\citenamefont {Andrade}, \citenamefont {Lorenzen}, \citenamefont {Marques},
		\citenamefont {Gross},\ and\ \citenamefont {Rubio}}]{octopus_local_field}%
	\BibitemOpen
	\bibfield  {author} {\bibinfo {author} {\bibfnamefont {A.}~\bibnamefont
			{Castro}}, \bibinfo {author} {\bibfnamefont {H.}~\bibnamefont {Appel}},
		\bibinfo {author} {\bibfnamefont {M.}~\bibnamefont {Oliveira}}, \bibinfo
		{author} {\bibfnamefont {C.~A.}\ \bibnamefont {Rozzi}}, \bibinfo {author}
		{\bibfnamefont {X.}~\bibnamefont {Andrade}}, \bibinfo {author} {\bibfnamefont
			{F.}~\bibnamefont {Lorenzen}}, \bibinfo {author} {\bibfnamefont {M.~A.~L.}\
			\bibnamefont {Marques}}, \bibinfo {author} {\bibfnamefont {E.~K.~U.}\
			\bibnamefont {Gross}},\ and\ \bibinfo {author} {\bibfnamefont
			{A.}~\bibnamefont {Rubio}},\ }\bibfield  {title} {\emph {\bibinfo {title}
			{octopus: a tool for the application of time‐dependent density functional
				theory}},\ }\href {https://doi.org/10.1002/pssb.200642067} {\bibfield
		{journal} {\bibinfo  {journal} {physica status solidi (b)}\ }\textbf
		{\bibinfo {volume} {243}},\ \bibinfo {pages} {2465--2488} (\bibinfo {year}
		{2006})}\BibitemShut {NoStop}%
	\bibitem [{\citenamefont {Chen}(2006)}]{cuprate_phase_diagram}%
	\BibitemOpen
	\bibfield  {author} {\bibinfo {author} {\bibfnamefont {C.-T.}\ \bibnamefont
			{Chen}},\ }\emph {\bibinfo {title} {Scanning {Tunneling} {Spectroscopy}
			{Studies} of {High}-{Temperature} {Cuprate} {Superconductors}}},\ \href
	{https://doi.org/10.1484/m.art-eb.4.00082} {Ph.D. thesis},\ \bibinfo
	{school} {California Institute of Technology} (\bibinfo {year}
	{2006})\BibitemShut {NoStop}%
	\bibitem [{\citenamefont {Tessman}\ \emph {et~al.}(1953)\citenamefont
		{Tessman}, \citenamefont {Kahn},\ and\ \citenamefont
		{Shockley}}]{O_polarizability}%
	\BibitemOpen
	\bibfield  {author} {\bibinfo {author} {\bibfnamefont {J.~R.}\ \bibnamefont
			{Tessman}}, \bibinfo {author} {\bibfnamefont {A.~H.}\ \bibnamefont {Kahn}},\
		and\ \bibinfo {author} {\bibfnamefont {W.}~\bibnamefont {Shockley}},\
	}\bibfield  {title} {\emph {\bibinfo {title} {Electronic {Polarizabilities}
				of {Ions} in {Crystals}}},\ }\href {https://doi.org/10.1103/PhysRev.92.890}
	{\bibfield  {journal} {\bibinfo  {journal} {Physical Review}\ }\textbf
		{\bibinfo {volume} {92}},\ \bibinfo {pages} {890--895} (\bibinfo {year}
		{1953})}\BibitemShut {NoStop}%
	\bibitem [{\citenamefont {Golden}\ \emph {et~al.}(2001)\citenamefont {Golden},
		\citenamefont {Dürr}, \citenamefont {Koitzsch}, \citenamefont {Legner},
		\citenamefont {Hu}, \citenamefont {Borisenko}, \citenamefont {Knupfer},\ and\
		\citenamefont {Fink}}]{apres_cuprate_electronic_structure}%
	\BibitemOpen
	\bibfield  {author} {\bibinfo {author} {\bibfnamefont {M.~S.}\ \bibnamefont
			{Golden}}, \bibinfo {author} {\bibfnamefont {C.}~\bibnamefont {Dürr}},
		\bibinfo {author} {\bibfnamefont {A.}~\bibnamefont {Koitzsch}}, \bibinfo
		{author} {\bibfnamefont {S.}~\bibnamefont {Legner}}, \bibinfo {author}
		{\bibfnamefont {Z.}~\bibnamefont {Hu}}, \bibinfo {author} {\bibfnamefont
			{S.}~\bibnamefont {Borisenko}}, \bibinfo {author} {\bibfnamefont
			{M.}~\bibnamefont {Knupfer}},\ and\ \bibinfo {author} {\bibfnamefont
			{J.}~\bibnamefont {Fink}},\ }\bibfield  {title} {\emph {\bibinfo {title} {The
				electronic structure of cuprates from high energy spectroscopy}},\ }\href
	{https://doi.org/10.1016/S0368-2048(01)00266-3} {\bibfield  {journal}
		{\bibinfo  {journal} {Journal of Electron Spectroscopy and Related
				Phenomena}\ }\bibinfo {series} {Strongly correlated systems},\ \textbf
		{\bibinfo {volume} {117-118}},\ \bibinfo {pages} {203--222} (\bibinfo {year}
		{2001})}\BibitemShut {NoStop}%
	\bibitem [{\citenamefont {Shirane}\ \emph {et~al.}(1987)\citenamefont
		{Shirane}, \citenamefont {Endoh}, \citenamefont {Birgeneau}, \citenamefont
		{Kastner}, \citenamefont {Hidaka}, \citenamefont {Oda}, \citenamefont
		{Suzuki},\ and\ \citenamefont {Murakami}}]{neutron_scattering_afm}%
	\BibitemOpen
	\bibfield  {author} {\bibinfo {author} {\bibfnamefont {G.}~\bibnamefont
			{Shirane}}, \bibinfo {author} {\bibfnamefont {Y.}~\bibnamefont {Endoh}},
		\bibinfo {author} {\bibfnamefont {R.~J.}\ \bibnamefont {Birgeneau}}, \bibinfo
		{author} {\bibfnamefont {M.~A.}\ \bibnamefont {Kastner}}, \bibinfo {author}
		{\bibfnamefont {Y.}~\bibnamefont {Hidaka}}, \bibinfo {author} {\bibfnamefont
			{M.}~\bibnamefont {Oda}}, \bibinfo {author} {\bibfnamefont {M.}~\bibnamefont
			{Suzuki}},\ and\ \bibinfo {author} {\bibfnamefont {T.}~\bibnamefont
			{Murakami}},\ }\bibfield  {title} {\emph {\bibinfo {title} {Two-dimensional
				antiferromagnetic quantum spin-fluid state in {La}$_{\textrm{2}}${CuO}}},\
	}\href {https://doi.org/10.1103/PhysRevLett.59.1613} {\bibfield  {journal}
		{\bibinfo  {journal} {Physical Review Letters}\ }\textbf {\bibinfo {volume}
			{59}},\ \bibinfo {pages} {1613--1616} (\bibinfo {year} {1987})}\BibitemShut
	{NoStop}%
	\bibitem [{\citenamefont {Eskes}\ \emph {et~al.}(1989)\citenamefont {Eskes},
		\citenamefont {Sawatzky},\ and\ \citenamefont {Feiner}}]{eskes_delta}%
	\BibitemOpen
	\bibfield  {author} {\bibinfo {author} {\bibfnamefont {H.}~\bibnamefont
			{Eskes}}, \bibinfo {author} {\bibfnamefont {G.}~\bibnamefont {Sawatzky}},\
		and\ \bibinfo {author} {\bibfnamefont {L.}~\bibnamefont {Feiner}},\
	}\bibfield  {title} {\emph {\bibinfo {title} {Effective transfer for singlets
				formed by hole doping in the high-{T}$_{\textrm{c}}$ superconductors}},\
	}\href {https://doi.org/10.1016/0921-4534(89)90415-2} {\bibfield  {journal}
		{\bibinfo  {journal} {Physica C: Superconductivity}\ }\textbf {\bibinfo
			{volume} {160}},\ \bibinfo {pages} {424--430} (\bibinfo {year}
		{1989})}\BibitemShut {NoStop}%
	\bibitem [{\citenamefont {Haase}\ \emph {et~al.}(2004)\citenamefont {Haase},
		\citenamefont {Sushkov}, \citenamefont {Horsch},\ and\ \citenamefont
		{Williams}}]{cuprate_nmr_hole_density}%
	\BibitemOpen
	\bibfield  {author} {\bibinfo {author} {\bibfnamefont {J.}~\bibnamefont
			{Haase}}, \bibinfo {author} {\bibfnamefont {O.~P.}\ \bibnamefont {Sushkov}},
		\bibinfo {author} {\bibfnamefont {P.}~\bibnamefont {Horsch}},\ and\ \bibinfo
		{author} {\bibfnamefont {G.~V.~M.}\ \bibnamefont {Williams}},\ }\bibfield
	{title} {\emph {\bibinfo {title} {Planar {Cu} and {O} hole densities in
				high-{T}$_{\textrm{c}}$ cuprates determined with {NMR}}},\ }\href
	{https://doi.org/10.1103/PhysRevB.69.094504} {\bibfield  {journal} {\bibinfo
			{journal} {Physical Review B}\ }\textbf {\bibinfo {volume} {69}},\ \bibinfo
		{pages} {094504} (\bibinfo {year} {2004})}\BibitemShut {NoStop}%
	\bibitem [{\citenamefont {Ogata}\ and\ \citenamefont
		{Fukuyama}(2008)}]{cuprate_review_params}%
	\BibitemOpen
	\bibfield  {author} {\bibinfo {author} {\bibfnamefont {M.}~\bibnamefont
			{Ogata}}\ and\ \bibinfo {author} {\bibfnamefont {H.}~\bibnamefont
			{Fukuyama}},\ }\bibfield  {title} {\emph {\bibinfo {title} {The t–{J} model
				for the oxide high- {T}$_{\textrm{c}}$ superconductors}},\ }\href
	{https://doi.org/10.1088/0034-4885/71/3/036501} {\bibfield  {journal}
		{\bibinfo  {journal} {Reports on Progress in Physics}\ }\textbf {\bibinfo
			{volume} {71}},\ \bibinfo {pages} {036501} (\bibinfo {year}
		{2008})}\BibitemShut {NoStop}%
	\bibitem [{\citenamefont {Wang}(1980)}]{exact_calcs_local_field}%
	\BibitemOpen
	\bibfield  {author} {\bibinfo {author} {\bibfnamefont {J.~C.}\ \bibnamefont
			{Wang}},\ }\bibfield  {title} {\emph {\bibinfo {title} {Local fields near a
				point-charge defect in cubic ionic crystals}},\ }\href
	{https://doi.org/10.1103/PhysRevB.22.2725} {\bibfield  {journal} {\bibinfo
			{journal} {Physical Review B}\ }\textbf {\bibinfo {volume} {22}},\ \bibinfo
		{pages} {2725--2730} (\bibinfo {year} {1980})}\BibitemShut {NoStop}%
	\bibitem [{\citenamefont {Weber}\ \emph {et~al.}(2012)\citenamefont {Weber},
		\citenamefont {Yee}, \citenamefont {Haule},\ and\ \citenamefont
		{Kotliar}}]{cuprate_tpp_tc_1}%
	\BibitemOpen
	\bibfield  {author} {\bibinfo {author} {\bibfnamefont {C.}~\bibnamefont
			{Weber}}, \bibinfo {author} {\bibfnamefont {C.}~\bibnamefont {Yee}}, \bibinfo
		{author} {\bibfnamefont {K.}~\bibnamefont {Haule}},\ and\ \bibinfo {author}
		{\bibfnamefont {G.}~\bibnamefont {Kotliar}},\ }\bibfield  {title} {\emph
		{\bibinfo {title} {Scaling of the transition temperature of hole-doped
				cuprate superconductors with the charge-transfer energy}},\ }\href
	{https://doi.org/10.1209/0295-5075/100/37001} {\bibfield  {journal} {\bibinfo
			{journal} {Europhysics Letters}\ }\textbf {\bibinfo {volume} {100}},\
		\bibinfo {pages} {37001} (\bibinfo {year} {2012})},\ \bibinfo {note}
	{publisher: EDP Sciences, IOP Publishing and Società Italiana di
		Fisica}\BibitemShut {NoStop}%
	\bibitem [{\citenamefont {Choi}\ \emph {et~al.}(2019)\citenamefont {Choi},
		\citenamefont {Di~Bernardo}, \citenamefont {Zhu}, \citenamefont {Lu},
		\citenamefont {Alpern}, \citenamefont {Zhang}, \citenamefont {Shapira},
		\citenamefont {Feighan}, \citenamefont {Sun}, \citenamefont {Robinson},
		\citenamefont {Paltiel}, \citenamefont {Millo}, \citenamefont {Wang},
		\citenamefont {Jia},\ and\ \citenamefont
		{MacManus-Driscoll}}]{cuprate_tdp_tc}%
	\BibitemOpen
	\bibfield  {author} {\bibinfo {author} {\bibfnamefont {E.-M.}\ \bibnamefont
			{Choi}}, \bibinfo {author} {\bibfnamefont {A.}~\bibnamefont {Di~Bernardo}},
		\bibinfo {author} {\bibfnamefont {B.}~\bibnamefont {Zhu}}, \bibinfo {author}
		{\bibfnamefont {P.}~\bibnamefont {Lu}}, \bibinfo {author} {\bibfnamefont
			{H.}~\bibnamefont {Alpern}}, \bibinfo {author} {\bibfnamefont {K.~H.~L.}\
			\bibnamefont {Zhang}}, \bibinfo {author} {\bibfnamefont {T.}~\bibnamefont
			{Shapira}}, \bibinfo {author} {\bibfnamefont {J.}~\bibnamefont {Feighan}},
		\bibinfo {author} {\bibfnamefont {X.}~\bibnamefont {Sun}}, \bibinfo {author}
		{\bibfnamefont {J.}~\bibnamefont {Robinson}}, \bibinfo {author}
		{\bibfnamefont {Y.}~\bibnamefont {Paltiel}}, \bibinfo {author} {\bibfnamefont
			{O.}~\bibnamefont {Millo}}, \bibinfo {author} {\bibfnamefont
			{H.}~\bibnamefont {Wang}}, \bibinfo {author} {\bibfnamefont {Q.}~\bibnamefont
			{Jia}},\ and\ \bibinfo {author} {\bibfnamefont {J.~L.}\ \bibnamefont
			{MacManus-Driscoll}},\ }\bibfield  {title} {\emph {\bibinfo {title} {{3D}
				strain-induced superconductivity in
				{La}$_{\textrm{2}}${CuO}$_{\textrm{4+{\textbackslash}delta}}$ using a simple
				vertically aligned nanocomposite approach}},\ }\href
	{https://doi.org/10.1126/sciadv.aav5532} {\bibfield  {journal} {\bibinfo
			{journal} {Science Advances}\ }\textbf {\bibinfo {volume} {5}},\ \bibinfo
		{pages} {eaav5532} (\bibinfo {year} {2019})},\ \bibinfo {note} {publisher:
		American Association for the Advancement of Science}\BibitemShut {NoStop}%
	\bibitem [{\citenamefont {Zegrodnik}\ \emph {et~al.}(2021)\citenamefont
		{Zegrodnik}, \citenamefont {Biborski}, \citenamefont {Fidrysiak},\ and\
		\citenamefont {Spałek}}]{cuprate_tpp_tc_2}%
	\BibitemOpen
	\bibfield  {author} {\bibinfo {author} {\bibfnamefont {M.}~\bibnamefont
			{Zegrodnik}}, \bibinfo {author} {\bibfnamefont {A.}~\bibnamefont {Biborski}},
		\bibinfo {author} {\bibfnamefont {M.}~\bibnamefont {Fidrysiak}},\ and\
		\bibinfo {author} {\bibfnamefont {J.}~\bibnamefont {Spałek}},\ }\bibfield
	{title} {\emph {\bibinfo {title} {Superconductivity in the three-band model
				of cuprates: nodal direction characteristics and influence of intersite
				interactions}},\ }\href {https://doi.org/10.1088/1361-648X/abcff6} {\bibfield
		{journal} {\bibinfo  {journal} {Journal of Physics: Condensed Matter}\
		}\textbf {\bibinfo {volume} {33}},\ \bibinfo {pages} {415601} (\bibinfo
		{year} {2021})},\ \bibinfo {note} {publisher: IOP Publishing}\BibitemShut
	{NoStop}%
	\bibitem [{\citenamefont {Keldysh}(1979)}]{low_dim_no_screening_1}%
	\BibitemOpen
	\bibfield  {author} {\bibinfo {author} {\bibfnamefont {L.~V.}\ \bibnamefont
			{Keldysh}},\ }\bibfield  {title} {\emph {\bibinfo {title} {Coulomb
				interaction in thin semiconductor and semimetal films}},\ }\href
	{https://www.worldscientific.com/doi/10.1142/9789811279461_0024} {\bibfield
		{journal} {\bibinfo  {journal} {Pis'ma Zh. Eksp. Teor. Fiz.}\ }\textbf
		{\bibinfo {volume} {29}},\ \bibinfo {pages} {716--719} (\bibinfo {year}
		{1979})}\BibitemShut {NoStop}%
	\bibitem [{\citenamefont {{E. A. Andryushin}}\ \emph
		{et~al.}(1980)\citenamefont {{E. A. Andryushin}}, \citenamefont {{L. V.
				Keldysh}}, \citenamefont {{V. A. Sanina}},\ and\ \citenamefont {{A. P.
				Silin}}}]{low_dim_no_screening_2}%
	\BibitemOpen
	\bibfield  {author} {\bibinfo {author} {\bibnamefont {{E. A. Andryushin}}},
		\bibinfo {author} {\bibnamefont {{L. V. Keldysh}}}, \bibinfo {author}
		{\bibnamefont {{V. A. Sanina}}},\ and\ \bibinfo {author} {\bibnamefont {{A.
					P. Silin}}},\ }\bibfield  {title} {\emph {\bibinfo {title} {Electron-hole
				liquid in thin semiconductor films}},\ }\href
	{http://www.jetp.ras.ru/cgi-bin/dn/e_052_04_0761.pdf} {\bibfield  {journal}
		{\bibinfo  {journal} {Zh. Eksp. Teor. Fiz.}\ }\textbf {\bibinfo {volume}
			{79}},\ \bibinfo {pages} {1509--1517} (\bibinfo {year} {1980})}\BibitemShut
	{NoStop}%
	\bibitem [{\citenamefont {Dell'Anna}\ and\ \citenamefont
		{Merano}(2016)}]{2d_clausius_dynamical}%
	\BibitemOpen
	\bibfield  {author} {\bibinfo {author} {\bibfnamefont {L.}~\bibnamefont
			{Dell'Anna}}\ and\ \bibinfo {author} {\bibfnamefont {M.}~\bibnamefont
			{Merano}},\ }\bibfield  {title} {\emph {\bibinfo {title} {Clausius-{Mossotti}
				{Lorentz}-{Lorenz} relations and retardation effects for two-dimensional
				crystals}},\ }\href {https://doi.org/10.1103/PhysRevA.93.053808} {\bibfield
		{journal} {\bibinfo  {journal} {Physical Review A}\ }\textbf {\bibinfo
			{volume} {93}},\ \bibinfo {pages} {1--7} (\bibinfo {year} {2016})},\ \bibinfo
	{note} {arXiv: 1602.05468}\BibitemShut {NoStop}%
	\bibitem [{\citenamefont {Peng}\ \emph {et~al.}(2017)\citenamefont {Peng},
		\citenamefont {Dellea}, \citenamefont {Minola}, \citenamefont {Conni},
		\citenamefont {Amorese}, \citenamefont {Di~Castro}, \citenamefont {De~Luca},
		\citenamefont {Kummer}, \citenamefont {Salluzzo}, \citenamefont {Sun},
		\citenamefont {Zhou}, \citenamefont {Balestrino}, \citenamefont {Le~Tacon},
		\citenamefont {Keimer}, \citenamefont {Braicovich}, \citenamefont {Brookes},\
		and\ \citenamefont {Ghiringhelli}}]{cuprate_apical_oxygen}%
	\BibitemOpen
	\bibfield  {author} {\bibinfo {author} {\bibfnamefont {Y.~Y.}\ \bibnamefont
			{Peng}}, \bibinfo {author} {\bibfnamefont {G.}~\bibnamefont {Dellea}},
		\bibinfo {author} {\bibfnamefont {M.}~\bibnamefont {Minola}}, \bibinfo
		{author} {\bibfnamefont {M.}~\bibnamefont {Conni}}, \bibinfo {author}
		{\bibfnamefont {A.}~\bibnamefont {Amorese}}, \bibinfo {author} {\bibfnamefont
			{D.}~\bibnamefont {Di~Castro}}, \bibinfo {author} {\bibfnamefont {G.~M.}\
			\bibnamefont {De~Luca}}, \bibinfo {author} {\bibfnamefont {K.}~\bibnamefont
			{Kummer}}, \bibinfo {author} {\bibfnamefont {M.}~\bibnamefont {Salluzzo}},
		\bibinfo {author} {\bibfnamefont {X.}~\bibnamefont {Sun}}, \bibinfo {author}
		{\bibfnamefont {X.~J.}\ \bibnamefont {Zhou}}, \bibinfo {author}
		{\bibfnamefont {G.}~\bibnamefont {Balestrino}}, \bibinfo {author}
		{\bibfnamefont {M.}~\bibnamefont {Le~Tacon}}, \bibinfo {author}
		{\bibfnamefont {B.}~\bibnamefont {Keimer}}, \bibinfo {author} {\bibfnamefont
			{L.}~\bibnamefont {Braicovich}}, \bibinfo {author} {\bibfnamefont {N.~B.}\
			\bibnamefont {Brookes}},\ and\ \bibinfo {author} {\bibfnamefont
			{G.}~\bibnamefont {Ghiringhelli}},\ }\bibfield  {title} {\emph {\bibinfo
			{title} {Influence of apical oxygen on the extent of in-plane exchange
				interaction in cuprate superconductors}},\ }\href
	{https://doi.org/10.1038/nphys4248} {\bibfield  {journal} {\bibinfo
			{journal} {Nature Physics}\ }\textbf {\bibinfo {volume} {13}},\ \bibinfo
		{pages} {1201--1206} (\bibinfo {year} {2017})},\ \bibinfo {note} {publisher:
		Nature Publishing Group}\BibitemShut {NoStop}%
	\bibitem [{\citenamefont {Hwang}(2021)}]{coherence_length}%
	\BibitemOpen
	\bibfield  {author} {\bibinfo {author} {\bibfnamefont {J.}~\bibnamefont
			{Hwang}},\ }\bibfield  {title} {\emph {\bibinfo {title} {Superconducting
				coherence length of hole-doped cuprates obtained from electron–boson
				spectral density function}},\ }\href
	{https://doi.org/10.1038/s41598-021-91163-w} {\bibfield  {journal} {\bibinfo
			{journal} {Scientific Reports}\ }\textbf {\bibinfo {volume} {11}},\ \bibinfo
		{pages} {11668} (\bibinfo {year} {2021})}\BibitemShut {NoStop}%
	\bibitem [{\citenamefont {Aji}\ \emph {et~al.}(2010)\citenamefont {Aji},
		\citenamefont {Shekhter},\ and\ \citenamefont
		{Varma}}]{varma_quantum_flucts}%
	\BibitemOpen
	\bibfield  {author} {\bibinfo {author} {\bibfnamefont {V.}~\bibnamefont
			{Aji}}, \bibinfo {author} {\bibfnamefont {A.}~\bibnamefont {Shekhter}},\ and\
		\bibinfo {author} {\bibfnamefont {C.~M.}\ \bibnamefont {Varma}},\ }\bibfield
	{title} {\emph {\bibinfo {title} {Theory of the coupling of quantum-critical
				fluctuations to fermions and d-wave superconductivity in cuprates}},\ }\href
	{https://doi.org/10.1103/PhysRevB.81.064515} {\bibfield  {journal} {\bibinfo
			{journal} {Physical Review B}\ }\textbf {\bibinfo {volume} {81}},\ \bibinfo
		{pages} {064515} (\bibinfo {year} {2010})}\BibitemShut {NoStop}%
	\bibitem [{\citenamefont {Varma}(1999)}]{varma_pseudogap}%
	\BibitemOpen
	\bibfield  {author} {\bibinfo {author} {\bibfnamefont {C.~M.}\ \bibnamefont
			{Varma}},\ }\bibfield  {title} {\emph {\bibinfo {title} {Pseudogap {Phase}
				and the {Quantum}-{Critical} {Point} in {Copper}-{Oxide} {Metals}}},\ }\href
	{https://doi.org/10.1103/PhysRevLett.83.3538} {\bibfield  {journal} {\bibinfo
			{journal} {Physical Review Letters}\ }\textbf {\bibinfo {volume} {83}},\
		\bibinfo {pages} {3538--3541} (\bibinfo {year} {1999})}\BibitemShut {NoStop}%
	\bibitem [{\citenamefont {Simon}\ and\ \citenamefont
		{Varma}(2002)}]{varma_pseudogap_experimental}%
	\BibitemOpen
	\bibfield  {author} {\bibinfo {author} {\bibfnamefont {M.~E.}\ \bibnamefont
			{Simon}}\ and\ \bibinfo {author} {\bibfnamefont {C.~M.}\ \bibnamefont
			{Varma}},\ }\bibfield  {title} {\emph {\bibinfo {title} {Detection and
				{Implications} of a {Time}-{Reversal} {Breaking} {State} in {Underdoped}
				{Cuprates}}},\ }\href {https://doi.org/10.1103/PhysRevLett.89.247003}
	{\bibfield  {journal} {\bibinfo  {journal} {Physical Review Letters}\
		}\textbf {\bibinfo {volume} {89}},\ \bibinfo {pages} {247003} (\bibinfo
		{year} {2002})}\BibitemShut {NoStop}%
	\bibitem [{\citenamefont {Loram}\ \emph {et~al.}(1993)\citenamefont {Loram},
		\citenamefont {Mirza}, \citenamefont {Cooper}, \citenamefont {Liang},\ and\
		\citenamefont {Wade}}]{pseudogap_cuprate_specific_heat}%
	\BibitemOpen
	\bibfield  {author} {\bibinfo {author} {\bibfnamefont {J.~W.}\ \bibnamefont
			{Loram}}, \bibinfo {author} {\bibfnamefont {K.~A.}\ \bibnamefont {Mirza}},
		\bibinfo {author} {\bibfnamefont {J.~R.}\ \bibnamefont {Cooper}}, \bibinfo
		{author} {\bibfnamefont {W.~Y.}\ \bibnamefont {Liang}},\ and\ \bibinfo
		{author} {\bibfnamefont {J.}~\bibnamefont {Wade}},\ }\bibfield  {title}
	{\emph {\bibinfo {title} {Electronic specific heat of
				{YBa}$_{\textrm{2}}${Cu}$_{\textrm{3}}${O}$_{\textrm{6+x}}$ from 1.8 to 300
				{K}}},\ }\href {https://doi.org/10.1103/PhysRevLett.71.1740} {\bibfield
		{journal} {\bibinfo  {journal} {Physical Review Letters}\ }\textbf {\bibinfo
			{volume} {71}},\ \bibinfo {pages} {1740--1743} (\bibinfo {year}
		{1993})}\BibitemShut {NoStop}%
	\bibitem [{\citenamefont {Alloul}\ \emph {et~al.}(1989)\citenamefont {Alloul},
		\citenamefont {Ohno},\ and\ \citenamefont {Mendels}}]{pseudogap_cuprate_nmr}%
	\BibitemOpen
	\bibfield  {author} {\bibinfo {author} {\bibfnamefont {H.}~\bibnamefont
			{Alloul}}, \bibinfo {author} {\bibfnamefont {T.}~\bibnamefont {Ohno}},\ and\
		\bibinfo {author} {\bibfnamefont {P.}~\bibnamefont {Mendels}},\ }\bibfield
	{title} {\emph {\bibinfo {title} {{89Y} {NMR} evidence for a fermi-liquid
				behavior in {YBa}$_{\textrm{2}}${Cu}$_{\textrm{3}}${O}$_{\textrm{6+x}}$}},\
	}\href {https://doi.org/10.1103/PhysRevLett.63.1700} {\bibfield  {journal}
		{\bibinfo  {journal} {Physical Review Letters}\ }\textbf {\bibinfo {volume}
			{63}},\ \bibinfo {pages} {1700--1703} (\bibinfo {year} {1989})}\BibitemShut
	{NoStop}%
	\bibitem [{\citenamefont {Shimojima}\ \emph {et~al.}(2014)\citenamefont
		{Shimojima}, \citenamefont {Sonobe}, \citenamefont {Malaeb}, \citenamefont
		{Shinada}, \citenamefont {Chainani}, \citenamefont {Shin}, \citenamefont
		{Yoshida}, \citenamefont {Ideta}, \citenamefont {Fujimori}, \citenamefont
		{Kumigashira}, \citenamefont {Ono}, \citenamefont {Nakashima}, \citenamefont
		{Anzai}, \citenamefont {Arita}, \citenamefont {Ino}, \citenamefont
		{Namatame}, \citenamefont {Taniguchi}, \citenamefont {Nakajima},
		\citenamefont {Uchida}, \citenamefont {Tomioka}, \citenamefont {Ito},
		\citenamefont {Kihou}, \citenamefont {Lee}, \citenamefont {Iyo},
		\citenamefont {Eisaki}, \citenamefont {Ohgushi}, \citenamefont {Kasahara},
		\citenamefont {Terashima}, \citenamefont {Ikeda}, \citenamefont {Shibauchi},
		\citenamefont {Matsuda},\ and\ \citenamefont
		{Ishizaka}}]{pseudogap_pnictides}%
	\BibitemOpen
	\bibfield  {author} {\bibinfo {author} {\bibfnamefont {T.}~\bibnamefont
			{Shimojima}}, \bibinfo {author} {\bibfnamefont {T.}~\bibnamefont {Sonobe}},
		\bibinfo {author} {\bibfnamefont {W.}~\bibnamefont {Malaeb}}, \bibinfo
		{author} {\bibfnamefont {K.}~\bibnamefont {Shinada}}, \bibinfo {author}
		{\bibfnamefont {A.}~\bibnamefont {Chainani}}, \bibinfo {author}
		{\bibfnamefont {S.}~\bibnamefont {Shin}}, \bibinfo {author} {\bibfnamefont
			{T.}~\bibnamefont {Yoshida}}, \bibinfo {author} {\bibfnamefont
			{S.}~\bibnamefont {Ideta}}, \bibinfo {author} {\bibfnamefont
			{A.}~\bibnamefont {Fujimori}}, \bibinfo {author} {\bibfnamefont
			{H.}~\bibnamefont {Kumigashira}}, \bibinfo {author} {\bibfnamefont
			{K.}~\bibnamefont {Ono}}, \bibinfo {author} {\bibfnamefont {Y.}~\bibnamefont
			{Nakashima}}, \bibinfo {author} {\bibfnamefont {H.}~\bibnamefont {Anzai}},
		\bibinfo {author} {\bibfnamefont {M.}~\bibnamefont {Arita}}, \bibinfo
		{author} {\bibfnamefont {A.}~\bibnamefont {Ino}}, \bibinfo {author}
		{\bibfnamefont {H.}~\bibnamefont {Namatame}}, \bibinfo {author}
		{\bibfnamefont {M.}~\bibnamefont {Taniguchi}}, \bibinfo {author}
		{\bibfnamefont {M.}~\bibnamefont {Nakajima}}, \bibinfo {author}
		{\bibfnamefont {S.}~\bibnamefont {Uchida}}, \bibinfo {author} {\bibfnamefont
			{Y.}~\bibnamefont {Tomioka}}, \bibinfo {author} {\bibfnamefont
			{T.}~\bibnamefont {Ito}}, \bibinfo {author} {\bibfnamefont {K.}~\bibnamefont
			{Kihou}}, \bibinfo {author} {\bibfnamefont {C.~H.}\ \bibnamefont {Lee}},
		\bibinfo {author} {\bibfnamefont {A.}~\bibnamefont {Iyo}}, \bibinfo {author}
		{\bibfnamefont {H.}~\bibnamefont {Eisaki}}, \bibinfo {author} {\bibfnamefont
			{K.}~\bibnamefont {Ohgushi}}, \bibinfo {author} {\bibfnamefont
			{S.}~\bibnamefont {Kasahara}}, \bibinfo {author} {\bibfnamefont
			{T.}~\bibnamefont {Terashima}}, \bibinfo {author} {\bibfnamefont
			{H.}~\bibnamefont {Ikeda}}, \bibinfo {author} {\bibfnamefont
			{T.}~\bibnamefont {Shibauchi}}, \bibinfo {author} {\bibfnamefont
			{Y.}~\bibnamefont {Matsuda}},\ and\ \bibinfo {author} {\bibfnamefont
			{K.}~\bibnamefont {Ishizaka}},\ }\bibfield  {title} {\emph {\bibinfo {title}
			{Pseudogap formation above the superconducting dome in iron pnictides}},\
	}\href {https://doi.org/10.1103/PhysRevB.89.045101} {\bibfield  {journal}
		{\bibinfo  {journal} {Physical Review B}\ }\textbf {\bibinfo {volume} {89}},\
		\bibinfo {pages} {045101} (\bibinfo {year} {2014})}\BibitemShut {NoStop}%
	\bibitem [{\citenamefont {Zhao}\ \emph {et~al.}(2021)\citenamefont {Zhao},
		\citenamefont {Zhou}, \citenamefont {Fu}, \citenamefont {Wang}, \citenamefont
		{Zhou}, \citenamefont {Cheng}, \citenamefont {Li}, \citenamefont {Song},
		\citenamefont {Li}, \citenamefont {Kang}, \citenamefont {Zheng},
		\citenamefont {Nie}, \citenamefont {Wu}, \citenamefont {Shan}, \citenamefont
		{Yu}, \citenamefont {Ying}, \citenamefont {Wang}, \citenamefont {Mei},
		\citenamefont {Wu},\ and\ \citenamefont {Chen}}]{pseudogap_nickelates}%
	\BibitemOpen
	\bibfield  {author} {\bibinfo {author} {\bibfnamefont {D.}~\bibnamefont
			{Zhao}}, \bibinfo {author} {\bibfnamefont {Y.}~\bibnamefont {Zhou}}, \bibinfo
		{author} {\bibfnamefont {Y.}~\bibnamefont {Fu}}, \bibinfo {author}
		{\bibfnamefont {L.}~\bibnamefont {Wang}}, \bibinfo {author} {\bibfnamefont
			{X.}~\bibnamefont {Zhou}}, \bibinfo {author} {\bibfnamefont {H.}~\bibnamefont
			{Cheng}}, \bibinfo {author} {\bibfnamefont {J.}~\bibnamefont {Li}}, \bibinfo
		{author} {\bibfnamefont {D.}~\bibnamefont {Song}}, \bibinfo {author}
		{\bibfnamefont {S.}~\bibnamefont {Li}}, \bibinfo {author} {\bibfnamefont
			{B.}~\bibnamefont {Kang}}, \bibinfo {author} {\bibfnamefont {L.}~\bibnamefont
			{Zheng}}, \bibinfo {author} {\bibfnamefont {L.}~\bibnamefont {Nie}}, \bibinfo
		{author} {\bibfnamefont {Z.}~\bibnamefont {Wu}}, \bibinfo {author}
		{\bibfnamefont {M.}~\bibnamefont {Shan}}, \bibinfo {author} {\bibfnamefont
			{F.}~\bibnamefont {Yu}}, \bibinfo {author} {\bibfnamefont {J.}~\bibnamefont
			{Ying}}, \bibinfo {author} {\bibfnamefont {S.}~\bibnamefont {Wang}}, \bibinfo
		{author} {\bibfnamefont {J.}~\bibnamefont {Mei}}, \bibinfo {author}
		{\bibfnamefont {T.}~\bibnamefont {Wu}},\ and\ \bibinfo {author}
		{\bibfnamefont {X.}~\bibnamefont {Chen}},\ }\bibfield  {title} {\emph
		{\bibinfo {title} {Intrinsic {Spin} {Susceptibility} and {Pseudogaplike}
				{Behavior} in {Infinite}-{Layer} {LaNiO}$_{\textrm{2}}$}},\ }\href
	{https://doi.org/10.1103/PhysRevLett.126.197001} {\bibfield  {journal}
		{\bibinfo  {journal} {Physical Review Letters}\ }\textbf {\bibinfo {volume}
			{126}},\ \bibinfo {pages} {197001} (\bibinfo {year} {2021})}\BibitemShut
	{NoStop}%
	\bibitem [{\citenamefont {Pracht}\ \emph {et~al.}(2016)\citenamefont {Pracht},
		\citenamefont {Bachar}, \citenamefont {Benfatto}, \citenamefont {Deutscher},
		\citenamefont {Farber}, \citenamefont {Dressel},\ and\ \citenamefont
		{Scheffler}}]{pseudogap_Al}%
	\BibitemOpen
	\bibfield  {author} {\bibinfo {author} {\bibfnamefont {U.~S.}\ \bibnamefont
			{Pracht}}, \bibinfo {author} {\bibfnamefont {N.}~\bibnamefont {Bachar}},
		\bibinfo {author} {\bibfnamefont {L.}~\bibnamefont {Benfatto}}, \bibinfo
		{author} {\bibfnamefont {G.}~\bibnamefont {Deutscher}}, \bibinfo {author}
		{\bibfnamefont {E.}~\bibnamefont {Farber}}, \bibinfo {author} {\bibfnamefont
			{M.}~\bibnamefont {Dressel}},\ and\ \bibinfo {author} {\bibfnamefont
			{M.}~\bibnamefont {Scheffler}},\ }\bibfield  {title} {\emph {\bibinfo {title}
			{Enhanced {Cooper} pairing versus suppressed phase coherence shaping the
				superconducting dome in coupled aluminum nanograins}},\ }\href
	{https://doi.org/10.1103/PhysRevB.93.100503} {\bibfield  {journal} {\bibinfo
			{journal} {Physical Review B}\ }\textbf {\bibinfo {volume} {93}},\ \bibinfo
		{pages} {100503} (\bibinfo {year} {2016})}\BibitemShut {NoStop}%
	\bibitem [{\citenamefont {Mondal}\ \emph {et~al.}(2011)\citenamefont {Mondal},
		\citenamefont {Kamlapure}, \citenamefont {Chand}, \citenamefont {Saraswat},
		\citenamefont {Kumar}, \citenamefont {Jesudasan}, \citenamefont {Benfatto},
		\citenamefont {Tripathi},\ and\ \citenamefont
		{Raychaudhuri}}]{pseudogap_NbN}%
	\BibitemOpen
	\bibfield  {author} {\bibinfo {author} {\bibfnamefont {M.}~\bibnamefont
			{Mondal}}, \bibinfo {author} {\bibfnamefont {A.}~\bibnamefont {Kamlapure}},
		\bibinfo {author} {\bibfnamefont {M.}~\bibnamefont {Chand}}, \bibinfo
		{author} {\bibfnamefont {G.}~\bibnamefont {Saraswat}}, \bibinfo {author}
		{\bibfnamefont {S.}~\bibnamefont {Kumar}}, \bibinfo {author} {\bibfnamefont
			{J.}~\bibnamefont {Jesudasan}}, \bibinfo {author} {\bibfnamefont
			{L.}~\bibnamefont {Benfatto}}, \bibinfo {author} {\bibfnamefont
			{V.}~\bibnamefont {Tripathi}},\ and\ \bibinfo {author} {\bibfnamefont
			{P.}~\bibnamefont {Raychaudhuri}},\ }\bibfield  {title} {\emph {\bibinfo
			{title} {Phase {Fluctuations} in a {Strongly} {Disordered} s-wave {NbN}
				{Superconductor} {Close} to the {Metal}-{Insulator} {Transition}}},\ }\href
	{https://doi.org/10.1103/PhysRevLett.106.047001} {\bibfield  {journal}
		{\bibinfo  {journal} {Physical Review Letters}\ }\textbf {\bibinfo {volume}
			{106}},\ \bibinfo {pages} {047001} (\bibinfo {year} {2011})}\BibitemShut
	{NoStop}%
	\bibitem [{\citenamefont {Sacépé}\ \emph {et~al.}(2010)\citenamefont
		{Sacépé}, \citenamefont {Chapelier}, \citenamefont {Baturina},
		\citenamefont {Vinokur}, \citenamefont {Baklanov},\ and\ \citenamefont
		{Sanquer}}]{pseudogap_TiN}%
	\BibitemOpen
	\bibfield  {author} {\bibinfo {author} {\bibfnamefont {B.}~\bibnamefont
			{Sacépé}}, \bibinfo {author} {\bibfnamefont {C.}~\bibnamefont {Chapelier}},
		\bibinfo {author} {\bibfnamefont {T.~I.}\ \bibnamefont {Baturina}}, \bibinfo
		{author} {\bibfnamefont {V.~M.}\ \bibnamefont {Vinokur}}, \bibinfo {author}
		{\bibfnamefont {M.~R.}\ \bibnamefont {Baklanov}},\ and\ \bibinfo {author}
		{\bibfnamefont {M.}~\bibnamefont {Sanquer}},\ }\bibfield  {title} {\emph
		{\bibinfo {title} {Pseudogap in a thin film of a conventional
				superconductor}},\ }\href {https://doi.org/10.1038/ncomms1140} {\bibfield
		{journal} {\bibinfo  {journal} {Nature Communications}\ }\textbf {\bibinfo
			{volume} {1}},\ \bibinfo {pages} {140} (\bibinfo {year} {2010})}\BibitemShut
	{NoStop}%
	\bibitem [{\citenamefont {Mannella}\ \emph {et~al.}(2005)\citenamefont
		{Mannella}, \citenamefont {Yang}, \citenamefont {Zhou}, \citenamefont
		{Zheng}, \citenamefont {Mitchell}, \citenamefont {Zaanen}, \citenamefont
		{Devereaux}, \citenamefont {Nagaosa}, \citenamefont {Hussain},\ and\
		\citenamefont {Shen}}]{pseudogap_manganite}%
	\BibitemOpen
	\bibfield  {author} {\bibinfo {author} {\bibfnamefont {N.}~\bibnamefont
			{Mannella}}, \bibinfo {author} {\bibfnamefont {W.~L.}\ \bibnamefont {Yang}},
		\bibinfo {author} {\bibfnamefont {X.~J.}\ \bibnamefont {Zhou}}, \bibinfo
		{author} {\bibfnamefont {H.}~\bibnamefont {Zheng}}, \bibinfo {author}
		{\bibfnamefont {J.~F.}\ \bibnamefont {Mitchell}}, \bibinfo {author}
		{\bibfnamefont {J.}~\bibnamefont {Zaanen}}, \bibinfo {author} {\bibfnamefont
			{T.~P.}\ \bibnamefont {Devereaux}}, \bibinfo {author} {\bibfnamefont
			{N.}~\bibnamefont {Nagaosa}}, \bibinfo {author} {\bibfnamefont
			{Z.}~\bibnamefont {Hussain}},\ and\ \bibinfo {author} {\bibfnamefont {Z.-X.}\
			\bibnamefont {Shen}},\ }\bibfield  {title} {\emph {\bibinfo {title} {Nodal
				quasiparticle in pseudogapped colossal magnetoresistive manganites}},\ }\href
	{https://doi.org/10.1038/nature04273} {\bibfield  {journal} {\bibinfo
			{journal} {Nature}\ }\textbf {\bibinfo {volume} {438}},\ \bibinfo {pages}
		{474--478} (\bibinfo {year} {2005})}\BibitemShut {NoStop}%
	\bibitem [{\citenamefont {van~den Brink}\ and\ \citenamefont {van
			Veenendaal}(2005)}]{rixs_correlation_function}%
	\BibitemOpen
	\bibfield  {author} {\bibinfo {author} {\bibfnamefont {J.}~\bibnamefont
			{van~den Brink}}\ and\ \bibinfo {author} {\bibfnamefont {M.}~\bibnamefont
			{van Veenendaal}},\ }\bibfield  {title} {\emph {\bibinfo {title} {Theory of
				indirect resonant inelastic {X}-ray scattering}},\ }\href
	{https://doi.org/10.1016/j.jpcs.2005.10.168} {\bibfield  {journal} {\bibinfo
			{journal} {Journal of Physics and Chemistry of Solids}\ }\bibinfo {series}
		{5th {International} {Conference} on {Inelastic} {X}-ray {Scattering} ({IXS}
			2004)},\ \textbf {\bibinfo {volume} {66}},\ \bibinfo {pages} {2145--2149}
		(\bibinfo {year} {2005})}\BibitemShut {NoStop}%
	\bibitem [{\citenamefont {Mukherjee}\ \emph {et~al.}(2020)\citenamefont
		{Mukherjee}, \citenamefont {Seo}, \citenamefont {Arik}, \citenamefont
		{Zhang}, \citenamefont {Zhang}, \citenamefont {Kirzhner}, \citenamefont
		{George}, \citenamefont {Markelz}, \citenamefont {Armitage}, \citenamefont
		{Koren}, \citenamefont {Wei},\ and\ \citenamefont
		{Cerne}}]{cuprate_linear_dichroism}%
	\BibitemOpen
	\bibfield  {author} {\bibinfo {author} {\bibfnamefont {A.}~\bibnamefont
			{Mukherjee}}, \bibinfo {author} {\bibfnamefont {J.}~\bibnamefont {Seo}},
		\bibinfo {author} {\bibfnamefont {M.~M.}\ \bibnamefont {Arik}}, \bibinfo
		{author} {\bibfnamefont {H.}~\bibnamefont {Zhang}}, \bibinfo {author}
		{\bibfnamefont {C.~C.}\ \bibnamefont {Zhang}}, \bibinfo {author}
		{\bibfnamefont {T.}~\bibnamefont {Kirzhner}}, \bibinfo {author}
		{\bibfnamefont {D.~K.}\ \bibnamefont {George}}, \bibinfo {author}
		{\bibfnamefont {A.~G.}\ \bibnamefont {Markelz}}, \bibinfo {author}
		{\bibfnamefont {N.~P.}\ \bibnamefont {Armitage}}, \bibinfo {author}
		{\bibfnamefont {G.}~\bibnamefont {Koren}}, \bibinfo {author} {\bibfnamefont
			{J.~Y.~T.}\ \bibnamefont {Wei}},\ and\ \bibinfo {author} {\bibfnamefont
			{J.}~\bibnamefont {Cerne}},\ }\bibfield  {title} {\emph {\bibinfo {title}
			{Linear dichroism infrared resonance in overdoped, underdoped, and optimally
				doped cuprate superconductors}},\ }\href
	{https://doi.org/10.1103/PhysRevB.102.054520} {\bibfield  {journal} {\bibinfo
			{journal} {Physical Review B}\ }\textbf {\bibinfo {volume} {102}},\ \bibinfo
		{pages} {054520} (\bibinfo {year} {2020})},\ \bibinfo {note} {publisher:
		American Physical Society}\BibitemShut {NoStop}%
	\bibitem [{\citenamefont {Iwasawa}\ \emph {et~al.}(2019)\citenamefont
		{Iwasawa}, \citenamefont {Dudin}, \citenamefont {Inui}, \citenamefont
		{Masui}, \citenamefont {Kim}, \citenamefont {Cacho},\ and\ \citenamefont
		{Hoesch}}]{YBCO_chains}%
	\BibitemOpen
	\bibfield  {author} {\bibinfo {author} {\bibfnamefont {H.}~\bibnamefont
			{Iwasawa}}, \bibinfo {author} {\bibfnamefont {P.}~\bibnamefont {Dudin}},
		\bibinfo {author} {\bibfnamefont {K.}~\bibnamefont {Inui}}, \bibinfo {author}
		{\bibfnamefont {T.}~\bibnamefont {Masui}}, \bibinfo {author} {\bibfnamefont
			{T.~K.}\ \bibnamefont {Kim}}, \bibinfo {author} {\bibfnamefont
			{C.}~\bibnamefont {Cacho}},\ and\ \bibinfo {author} {\bibfnamefont
			{M.}~\bibnamefont {Hoesch}},\ }\bibfield  {title} {\emph {\bibinfo {title}
			{Buried double {CuO} chains in
				{YBa}$_{\textrm{2}}${Cu}$_{\textrm{4}}${O}$_{\textrm{8}}$ uncovered by
				nano-{ARPES}}},\ }\href {https://doi.org/10.1103/PhysRevB.99.140510}
	{\bibfield  {journal} {\bibinfo  {journal} {Physical Review B}\ }\textbf
		{\bibinfo {volume} {99}},\ \bibinfo {pages} {140510} (\bibinfo {year}
		{2019})},\ \bibinfo {note} {publisher: American Physical Society}\BibitemShut
	{NoStop}%
	\bibitem [{\citenamefont {Sjöstrand}\ \emph {et~al.}(2019)\citenamefont
		{Sjöstrand}, \citenamefont {Nilsson}, \citenamefont {Friedrich},\ and\
		\citenamefont {Aryasetiawan}}]{cuprate_U_W_real_space}%
	\BibitemOpen
	\bibfield  {author} {\bibinfo {author} {\bibfnamefont {T.~J.}\ \bibnamefont
			{Sjöstrand}}, \bibinfo {author} {\bibfnamefont {F.}~\bibnamefont {Nilsson}},
		\bibinfo {author} {\bibfnamefont {C.}~\bibnamefont {Friedrich}},\ and\
		\bibinfo {author} {\bibfnamefont {F.}~\bibnamefont {Aryasetiawan}},\
	}\bibfield  {title} {\emph {\bibinfo {title} {Position representation of
				effective electron-electron interactions in solids}},\ }\href
	{https://doi.org/10.1103/PhysRevB.99.195136} {\bibfield  {journal} {\bibinfo
			{journal} {Physical Review B}\ }\textbf {\bibinfo {volume} {99}},\ \bibinfo
		{pages} {195136} (\bibinfo {year} {2019})},\ \bibinfo {note} {publisher:
		American Physical Society}\BibitemShut {NoStop}%
	\bibitem [{\citenamefont {Eliashberg}(1960)}]{eliashberg_original}%
	\BibitemOpen
	\bibfield  {author} {\bibinfo {author} {\bibfnamefont {G.}~\bibnamefont
			{Eliashberg}},\ }\bibfield  {title} {\emph {\bibinfo {title} {Interactions
				between electrons and lattice vibrations in a superconductor}},\ }\href
	{https://www.osti.gov/biblio/7354388} {\bibfield  {journal} {\bibinfo
			{journal} {Sov. Phys. - JETP}\ }\textbf {\bibinfo {volume} {11:3}} (\bibinfo
		{year} {1960})}\BibitemShut {NoStop}%
	\bibitem [{\citenamefont {Allen}\ and\ \citenamefont
		{Dynes}(1975)}]{dynes_original}%
	\BibitemOpen
	\bibfield  {author} {\bibinfo {author} {\bibfnamefont {P.~B.}\ \bibnamefont
			{Allen}}\ and\ \bibinfo {author} {\bibfnamefont {R.~C.}\ \bibnamefont
			{Dynes}},\ }\bibfield  {title} {\emph {\bibinfo {title} {Transition
				temperature of strong-coupled superconductors reanalyzed}},\ }\href
	{https://doi.org/10.1103/PhysRevB.12.905} {\bibfield  {journal} {\bibinfo
			{journal} {Physical Review B}\ }\textbf {\bibinfo {volume} {12}},\ \bibinfo
		{pages} {905--922} (\bibinfo {year} {1975})}\BibitemShut {NoStop}%
	\bibitem [{\citenamefont {Fauqué}\ \emph {et~al.}(2024)\citenamefont
		{Fauqué}, \citenamefont {Jiang}, \citenamefont {Fennell}, \citenamefont
		{Roessli}, \citenamefont {Ivanov}, \citenamefont {Roux-Byl}, \citenamefont
		{Baptiste}, \citenamefont {Bourges}, \citenamefont {Behnia},\ and\
		\citenamefont {Tomioka}}]{polarization_length_scale_doping}%
	\BibitemOpen
	\bibfield  {author} {\bibinfo {author} {\bibfnamefont {B.}~\bibnamefont
			{Fauqué}}, \bibinfo {author} {\bibfnamefont {S.}~\bibnamefont {Jiang}},
		\bibinfo {author} {\bibfnamefont {T.}~\bibnamefont {Fennell}}, \bibinfo
		{author} {\bibfnamefont {B.}~\bibnamefont {Roessli}}, \bibinfo {author}
		{\bibfnamefont {A.}~\bibnamefont {Ivanov}}, \bibinfo {author} {\bibfnamefont
			{C.}~\bibnamefont {Roux-Byl}}, \bibinfo {author} {\bibfnamefont
			{B.}~\bibnamefont {Baptiste}}, \bibinfo {author} {\bibfnamefont
			{P.}~\bibnamefont {Bourges}}, \bibinfo {author} {\bibfnamefont
			{K.}~\bibnamefont {Behnia}},\ and\ \bibinfo {author} {\bibfnamefont
			{Y.}~\bibnamefont {Tomioka}},\ }\bibfield  {title} {\emph {\bibinfo {title}
			{The polarisation fluctuation length scale shaping the superconducting dome
				of {SrTiO}$_{\textrm{3}}$}},\ }\bibfield  {journal} {\bibinfo  {journal}
		{arXiv}\ }\textbf {\bibinfo {volume} {2404.04154}},\ \href
	{https://doi.org/10.48550/arXiv.2404.04154} {10.48550/arXiv.2404.04154}
	(\bibinfo {year} {2024}),\ \bibinfo {note} {arXiv:2404.04154
		[cond-mat]}\BibitemShut {NoStop}%
\end{thebibliography}

\begin{thebibliography}{15}%
	\makeatletter
	\providecommand \@ifxundefined [1]{%
		\@ifx{#1\undefined}
	}%
	\providecommand \@ifnum [1]{%
		\ifnum #1\expandafter \@firstoftwo
		\else \expandafter \@secondoftwo
		\fi
	}%
	\providecommand \@ifx [1]{%
		\ifx #1\expandafter \@firstoftwo
		\else \expandafter \@secondoftwo
		\fi
	}%
	\providecommand \natexlab [1]{#1}%
	\providecommand \enquote  [1]{``#1''}%
	\providecommand \bibnamefont  [1]{#1}%
	\providecommand \bibfnamefont [1]{#1}%
	\providecommand \citenamefont [1]{#1}%
	\providecommand \href@noop [0]{\@secondoftwo}%
	\providecommand \href [0]{\begingroup \@sanitize@url \@href}%
	\providecommand \@href[1]{\@@startlink{#1}\@@href}%
	\providecommand \@@href[1]{\endgroup#1\@@endlink}%
	\providecommand \@sanitize@url [0]{\catcode `\\12\catcode `\$12\catcode
		`\&12\catcode `\#12\catcode `\^12\catcode `\_12\catcode `\%12\relax}%
	\providecommand \@@startlink[1]{}%
	\providecommand \@@endlink[0]{}%
	\providecommand \url  [0]{\begingroup\@sanitize@url \@url }%
	\providecommand \@url [1]{\endgroup\@href {#1}{\urlprefix }}%
	\providecommand \urlprefix  [0]{URL }%
	\providecommand \Eprint [0]{\href }%
	\providecommand \doibase [0]{https://doi.org/}%
	\providecommand \selectlanguage [0]{\@gobble}%
	\providecommand \bibinfo  [0]{\@secondoftwo}%
	\providecommand \bibfield  [0]{\@secondoftwo}%
	\providecommand \translation [1]{[#1]}%
	\providecommand \BibitemOpen [0]{}%
	\providecommand \bibitemStop [0]{}%
	\providecommand \bibitemNoStop [0]{.\EOS\space}%
	\providecommand \EOS [0]{\spacefactor3000\relax}%
	\providecommand \BibitemShut  [1]{\csname bibitem#1\endcsname}%
	\let\auto@bib@innerbib\@empty
	\bibitem [{\citenamefont {Macridin}\ \emph {et~al.}(2005)\citenamefont
		{Macridin}, \citenamefont {Jarrell}, \citenamefont {Maier},\ and\
		\citenamefont {Sawatzky}}]{macridin_charge_density}%
	\BibitemOpen
	\bibfield  {author} {\bibinfo {author} {\bibfnamefont {A.}~\bibnamefont
			{Macridin}}, \bibinfo {author} {\bibfnamefont {M.}~\bibnamefont {Jarrell}},
		\bibinfo {author} {\bibfnamefont {T.}~\bibnamefont {Maier}},\ and\ \bibinfo
		{author} {\bibfnamefont {G.~A.}\ \bibnamefont {Sawatzky}},\ }\bibfield
	{title} {\emph {\bibinfo {title} {Physics of cuprates with the two-band
				{Hubbard} model: {The} validity of the one-band {Hubbard} model}},\
	}\bibfield  {journal} {\bibinfo  {journal} {Physical Review B}\ }\textbf
	{\bibinfo {volume} {71}},\ \href {https://doi.org/10.1103/PhysRevB.71.134527}
	{10.1103/PhysRevB.71.134527} (\bibinfo {year} {2005}),\ \bibinfo {note}
	{arXiv: cond-mat/0411092}\BibitemShut {NoStop}%
	\bibitem [{\citenamefont {Kelly}\ and\ \citenamefont
		{Palumbo}(1973)}]{O_ionization_potential}%
	\BibitemOpen
	\bibfield  {author} {\bibinfo {author} {\bibfnamefont {R.~L.}\ \bibnamefont
			{Kelly}}\ and\ \bibinfo {author} {\bibfnamefont {L.~J.}\ \bibnamefont
			{Palumbo}},\ }\href {https://www.osti.gov/biblio/4332843} {\emph {\bibinfo
			{title} {Atomic and ionic emission lines below 2000 angstroms: hydrogen
				through krypton}}},\ \bibinfo {type} {Tech. {Rep}.}\ \bibinfo {number}
	{NRL-7599}\ (\bibinfo  {institution} {Naval Research Lab., Washington, D.C.
		(USA)},\ \bibinfo {year} {1973})\BibitemShut {NoStop}%
	\bibitem [{\citenamefont {Kristiansson}\ \emph {et~al.}(2022)\citenamefont
		{Kristiansson}, \citenamefont {Chartkunchand}, \citenamefont {Eklund},
		\citenamefont {Hole}, \citenamefont {Anderson}, \citenamefont {de~Ruette},
		\citenamefont {Kamińska}, \citenamefont {Punnakayathil}, \citenamefont
		{Navarro-Navarrete}, \citenamefont {Sigurdsson}, \citenamefont {Grumer},
		\citenamefont {Simonsson}, \citenamefont {Björkhage}, \citenamefont
		{Rosén}, \citenamefont {Reinhed}, \citenamefont {Blom}, \citenamefont
		{Källberg}, \citenamefont {Alexander}, \citenamefont {Cederquist},
		\citenamefont {Zettergren}, \citenamefont {Schmidt},\ and\ \citenamefont
		{Hanstorp}}]{O_electron_affinity}%
	\BibitemOpen
	\bibfield  {author} {\bibinfo {author} {\bibfnamefont {M.~K.}\ \bibnamefont
			{Kristiansson}}, \bibinfo {author} {\bibfnamefont {K.}~\bibnamefont
			{Chartkunchand}}, \bibinfo {author} {\bibfnamefont {G.}~\bibnamefont
			{Eklund}}, \bibinfo {author} {\bibfnamefont {O.~M.}\ \bibnamefont {Hole}},
		\bibinfo {author} {\bibfnamefont {E.~K.}\ \bibnamefont {Anderson}}, \bibinfo
		{author} {\bibfnamefont {N.}~\bibnamefont {de~Ruette}}, \bibinfo {author}
		{\bibfnamefont {M.}~\bibnamefont {Kamińska}}, \bibinfo {author}
		{\bibfnamefont {N.}~\bibnamefont {Punnakayathil}}, \bibinfo {author}
		{\bibfnamefont {J.~E.}\ \bibnamefont {Navarro-Navarrete}}, \bibinfo {author}
		{\bibfnamefont {S.}~\bibnamefont {Sigurdsson}}, \bibinfo {author}
		{\bibfnamefont {J.}~\bibnamefont {Grumer}}, \bibinfo {author} {\bibfnamefont
			{A.}~\bibnamefont {Simonsson}}, \bibinfo {author} {\bibfnamefont
			{M.}~\bibnamefont {Björkhage}}, \bibinfo {author} {\bibfnamefont
			{S.}~\bibnamefont {Rosén}}, \bibinfo {author} {\bibfnamefont
			{P.}~\bibnamefont {Reinhed}}, \bibinfo {author} {\bibfnamefont
			{M.}~\bibnamefont {Blom}}, \bibinfo {author} {\bibfnamefont {A.}~\bibnamefont
			{Källberg}}, \bibinfo {author} {\bibfnamefont {J.~D.}\ \bibnamefont
			{Alexander}}, \bibinfo {author} {\bibfnamefont {H.}~\bibnamefont
			{Cederquist}}, \bibinfo {author} {\bibfnamefont {H.}~\bibnamefont
			{Zettergren}}, \bibinfo {author} {\bibfnamefont {H.~T.}\ \bibnamefont
			{Schmidt}},\ and\ \bibinfo {author} {\bibfnamefont {D.}~\bibnamefont
			{Hanstorp}},\ }\bibfield  {title} {\emph {\bibinfo {title} {High-precision
				electron affinity of oxygen}},\ }\href
	{https://doi.org/10.1038/s41467-022-33438-y} {\bibfield  {journal} {\bibinfo
			{journal} {Nature Communications}\ }\textbf {\bibinfo {volume} {13}},\
		\bibinfo {pages} {5906} (\bibinfo {year} {2022})}\BibitemShut {NoStop}%
	\bibitem [{\citenamefont {James E.~Huheey}(1993)}]{O_second_electron_affinity}%
	\BibitemOpen
	\bibfield  {author} {\bibinfo {author} {\bibfnamefont {R.~L.~K.}\
			\bibnamefont {James E.~Huheey}, \bibfnamefont {Ellen A.~Keiter}},\ }\href
	{https://www.semanticscholar.org/paper/Inorganic-chemistry%3B-principles-of-structure-and-Huheey/c0f189ba30e082b6ddfa98dc03dbd1d17d39f1b6}
	{\emph {\bibinfo {title} {Inorganic {Chemistry} {Principles} of {Structure}
				and {Relativity}}}}\ (\bibinfo  {publisher} {HarperCollins},\ \bibinfo
	{address} {New York},\ \bibinfo {year} {1993})\BibitemShut {NoStop}%
	\bibitem [{\citenamefont {Sugar}\ and\ \citenamefont
		{Musgrove}(1990)}]{Cu_ionization_potential}%
	\BibitemOpen
	\bibfield  {author} {\bibinfo {author} {\bibfnamefont {J.}~\bibnamefont
			{Sugar}}\ and\ \bibinfo {author} {\bibfnamefont {A.}~\bibnamefont
			{Musgrove}},\ }\bibfield  {title} {\emph {\bibinfo {title} {Energy {Levels}
				of {Copper}, {Cu} {I} through {Cu} {XXIX}}},\ }\href
	{https://doi.org/10.1063/1.555855} {\bibfield  {journal} {\bibinfo  {journal}
			{Journal of Physical and Chemical Reference Data}\ }\textbf {\bibinfo
			{volume} {19}},\ \bibinfo {pages} {527--616} (\bibinfo {year}
		{1990})}\BibitemShut {NoStop}%
	\bibitem [{\citenamefont {Kramida}\ \emph {et~al.}(2017)\citenamefont
		{Kramida}, \citenamefont {Nave},\ and\ \citenamefont
		{Reader}}]{Cu_second_ionization_energy}%
	\BibitemOpen
	\bibfield  {author} {\bibinfo {author} {\bibfnamefont {A.}~\bibnamefont
			{Kramida}}, \bibinfo {author} {\bibfnamefont {G.}~\bibnamefont {Nave}},\ and\
		\bibinfo {author} {\bibfnamefont {J.}~\bibnamefont {Reader}},\ }\bibfield
	{title} {\emph {\bibinfo {title} {The {Cu} {II} {Spectrum}}},\ }\href
	{https://doi.org/10.3390/atoms5010009} {\bibfield  {journal} {\bibinfo
			{journal} {Atoms}\ }\textbf {\bibinfo {volume} {5}},\ \bibinfo {pages} {9}
		(\bibinfo {year} {2017})}\BibitemShut {NoStop}%
	\bibitem [{\citenamefont {Virtanen}\ \emph {et~al.}(2020)\citenamefont
		{Virtanen}, \citenamefont {Gommers}, \citenamefont {Oliphant}, \citenamefont
		{Haberland}, \citenamefont {Reddy}, \citenamefont {Cournapeau}, \citenamefont
		{Burovski}, \citenamefont {Peterson}, \citenamefont {Weckesser},
		\citenamefont {Bright}, \citenamefont {van~der Walt}, \citenamefont {Brett},
		\citenamefont {Wilson}, \citenamefont {Millman}, \citenamefont {Mayorov},
		\citenamefont {Nelson}, \citenamefont {Jones}, \citenamefont {Kern},
		\citenamefont {Larson}, \citenamefont {Carey}, \citenamefont {Polat},
		\citenamefont {Feng}, \citenamefont {Moore}, \citenamefont {VanderPlas},
		\citenamefont {Laxalde}, \citenamefont {Perktold}, \citenamefont {Cimrman},
		\citenamefont {Henriksen}, \citenamefont {Quintero}, \citenamefont {Harris},
		\citenamefont {Archibald}, \citenamefont {Ribeiro}, \citenamefont
		{Pedregosa},\ and\ \citenamefont {van Mulbregt}}]{SciPy}%
	\BibitemOpen
	\bibfield  {author} {\bibinfo {author} {\bibfnamefont {P.}~\bibnamefont
			{Virtanen}}, \bibinfo {author} {\bibfnamefont {R.}~\bibnamefont {Gommers}},
		\bibinfo {author} {\bibfnamefont {T.~E.}\ \bibnamefont {Oliphant}}, \bibinfo
		{author} {\bibfnamefont {M.}~\bibnamefont {Haberland}}, \bibinfo {author}
		{\bibfnamefont {T.}~\bibnamefont {Reddy}}, \bibinfo {author} {\bibfnamefont
			{D.}~\bibnamefont {Cournapeau}}, \bibinfo {author} {\bibfnamefont
			{E.}~\bibnamefont {Burovski}}, \bibinfo {author} {\bibfnamefont
			{P.}~\bibnamefont {Peterson}}, \bibinfo {author} {\bibfnamefont
			{W.}~\bibnamefont {Weckesser}}, \bibinfo {author} {\bibfnamefont
			{J.}~\bibnamefont {Bright}}, \bibinfo {author} {\bibfnamefont {S.~J.}\
			\bibnamefont {van~der Walt}}, \bibinfo {author} {\bibfnamefont
			{M.}~\bibnamefont {Brett}}, \bibinfo {author} {\bibfnamefont
			{J.}~\bibnamefont {Wilson}}, \bibinfo {author} {\bibfnamefont {K.~J.}\
			\bibnamefont {Millman}}, \bibinfo {author} {\bibfnamefont {N.}~\bibnamefont
			{Mayorov}}, \bibinfo {author} {\bibfnamefont {A.~R.~J.}\ \bibnamefont
			{Nelson}}, \bibinfo {author} {\bibfnamefont {E.}~\bibnamefont {Jones}},
		\bibinfo {author} {\bibfnamefont {R.}~\bibnamefont {Kern}}, \bibinfo {author}
		{\bibfnamefont {E.}~\bibnamefont {Larson}}, \bibinfo {author} {\bibfnamefont
			{C.~J.}\ \bibnamefont {Carey}}, \bibinfo {author} {\bibfnamefont
			{I.}~\bibnamefont {Polat}}, \bibinfo {author} {\bibfnamefont
			{Y.}~\bibnamefont {Feng}}, \bibinfo {author} {\bibfnamefont {E.~W.}\
			\bibnamefont {Moore}}, \bibinfo {author} {\bibfnamefont {J.}~\bibnamefont
			{VanderPlas}}, \bibinfo {author} {\bibfnamefont {D.}~\bibnamefont {Laxalde}},
		\bibinfo {author} {\bibfnamefont {J.}~\bibnamefont {Perktold}}, \bibinfo
		{author} {\bibfnamefont {R.}~\bibnamefont {Cimrman}}, \bibinfo {author}
		{\bibfnamefont {I.}~\bibnamefont {Henriksen}}, \bibinfo {author}
		{\bibfnamefont {E.~A.}\ \bibnamefont {Quintero}}, \bibinfo {author}
		{\bibfnamefont {C.~R.}\ \bibnamefont {Harris}}, \bibinfo {author}
		{\bibfnamefont {A.~M.}\ \bibnamefont {Archibald}}, \bibinfo {author}
		{\bibfnamefont {A.~H.}\ \bibnamefont {Ribeiro}}, \bibinfo {author}
		{\bibfnamefont {F.}~\bibnamefont {Pedregosa}},\ and\ \bibinfo {author}
		{\bibfnamefont {P.}~\bibnamefont {van Mulbregt}},\ }\bibfield  {title} {\emph
		{\bibinfo {title} {{SciPy} 1.0: fundamental algorithms for scientific
				computing in {Python}}},\ }\href {https://doi.org/10.1038/s41592-019-0686-2}
	{\bibfield  {journal} {\bibinfo  {journal} {Nature Methods}\ }\textbf
		{\bibinfo {volume} {17}},\ \bibinfo {pages} {261--272} (\bibinfo {year}
		{2020})},\ \bibinfo {note} {publisher: Nature Publishing Group}\BibitemShut
	{NoStop}%
	\bibitem [{\citenamefont {Ogata}\ and\ \citenamefont
		{Fukuyama}(2008)}]{cuprate_review_params}%
	\BibitemOpen
	\bibfield  {author} {\bibinfo {author} {\bibfnamefont {M.}~\bibnamefont
			{Ogata}}\ and\ \bibinfo {author} {\bibfnamefont {H.}~\bibnamefont
			{Fukuyama}},\ }\bibfield  {title} {\emph {\bibinfo {title} {The t–{J} model
				for the oxide high- {T}$_{\textrm{c}}$ superconductors}},\ }\href
	{https://doi.org/10.1088/0034-4885/71/3/036501} {\bibfield  {journal}
		{\bibinfo  {journal} {Reports on Progress in Physics}\ }\textbf {\bibinfo
			{volume} {71}},\ \bibinfo {pages} {036501} (\bibinfo {year}
		{2008})}\BibitemShut {NoStop}%
	\bibitem [{\citenamefont {McMahan}\ \emph {et~al.}(1990)\citenamefont
		{McMahan}, \citenamefont {Annett},\ and\ \citenamefont
		{Martin}}]{cuprate_dft_params}%
	\BibitemOpen
	\bibfield  {author} {\bibinfo {author} {\bibfnamefont {A.~K.}\ \bibnamefont
			{McMahan}}, \bibinfo {author} {\bibfnamefont {J.~F.}\ \bibnamefont
			{Annett}},\ and\ \bibinfo {author} {\bibfnamefont {R.~M.}\ \bibnamefont
			{Martin}},\ }\bibfield  {title} {\emph {\bibinfo {title} {Cuprate parameters
				from numerical {Wannier} functions}},\ }\href
	{https://doi.org/10.1103/PhysRevB.42.6268} {\bibfield  {journal} {\bibinfo
			{journal} {Physical Review B}\ }\textbf {\bibinfo {volume} {42}},\ \bibinfo
		{pages} {6268--6282} (\bibinfo {year} {1990})}\BibitemShut {NoStop}%
	\bibitem [{\citenamefont {Zhang}\ and\ \citenamefont
		{Rice}(1988)}]{zr_singlet}%
	\BibitemOpen
	\bibfield  {author} {\bibinfo {author} {\bibfnamefont {F.~C.}\ \bibnamefont
			{Zhang}}\ and\ \bibinfo {author} {\bibfnamefont {T.~M.}\ \bibnamefont
			{Rice}},\ }\bibfield  {title} {\emph {\bibinfo {title} {Effective
				{Hamiltonian} for the superconducting {Cu} oxides}},\ }\href
	{https://doi.org/10.1103/PhysRevB.37.3759} {\bibfield  {journal} {\bibinfo
			{journal} {Physical Review B}\ }\textbf {\bibinfo {volume} {37}},\ \bibinfo
		{pages} {3759--3761} (\bibinfo {year} {1988})}\BibitemShut {NoStop}%
	\bibitem [{\citenamefont {Eskes}\ \emph {et~al.}(1989)\citenamefont {Eskes},
		\citenamefont {Sawatzky},\ and\ \citenamefont {Feiner}}]{eskes_delta}%
	\BibitemOpen
	\bibfield  {author} {\bibinfo {author} {\bibfnamefont {H.}~\bibnamefont
			{Eskes}}, \bibinfo {author} {\bibfnamefont {G.}~\bibnamefont {Sawatzky}},\
		and\ \bibinfo {author} {\bibfnamefont {L.}~\bibnamefont {Feiner}},\
	}\bibfield  {title} {\emph {\bibinfo {title} {Effective transfer for singlets
				formed by hole doping in the high-{T}$_{\textrm{c}}$ superconductors}},\
	}\href {https://doi.org/10.1016/0921-4534(89)90415-2} {\bibfield  {journal}
		{\bibinfo  {journal} {Physica C: Superconductivity}\ }\textbf {\bibinfo
			{volume} {160}},\ \bibinfo {pages} {424--430} (\bibinfo {year}
		{1989})}\BibitemShut {NoStop}%
	\bibitem [{\citenamefont {Tsukada}\ \emph {et~al.}(2006)\citenamefont
		{Tsukada}, \citenamefont {Shibata}, \citenamefont {Noda}, \citenamefont
		{Yamamoto},\ and\ \citenamefont {Naito}}]{charge_transfer_madelung_vacuum}%
	\BibitemOpen
	\bibfield  {author} {\bibinfo {author} {\bibfnamefont {A.}~\bibnamefont
			{Tsukada}}, \bibinfo {author} {\bibfnamefont {H.}~\bibnamefont {Shibata}},
		\bibinfo {author} {\bibfnamefont {M.}~\bibnamefont {Noda}}, \bibinfo {author}
		{\bibfnamefont {H.}~\bibnamefont {Yamamoto}},\ and\ \bibinfo {author}
		{\bibfnamefont {M.}~\bibnamefont {Naito}},\ }\bibfield  {title} {\emph
		{\bibinfo {title} {Charge transfer gap for
				{T}'-{RE}$_{\textrm{2}}${CuO}$_{\textrm{4}}$ and
				{T}-{La}$_{\textrm{2}}${CuO}$_{\textrm{4}}$ as estimated from {Madelung}
				potential calculations}},\ }\href
	{https://doi.org/10.1016/j.physc.2006.03.087} {\bibfield  {journal} {\bibinfo
			{journal} {Physica C: Superconductivity and its Applications}\ }\textbf
		{\bibinfo {volume} {445-448}},\ \bibinfo {pages} {94--96} (\bibinfo {year}
		{2006})}\BibitemShut {NoStop}%
	\bibitem [{\citenamefont {Stechel}\ and\ \citenamefont
		{Jennison}(1988)}]{cuprate_experimental_delta_1.5}%
	\BibitemOpen
	\bibfield  {author} {\bibinfo {author} {\bibfnamefont {E.~B.}\ \bibnamefont
			{Stechel}}\ and\ \bibinfo {author} {\bibfnamefont {D.~R.}\ \bibnamefont
			{Jennison}},\ }\bibfield  {title} {\emph {\bibinfo {title} {Electronic
				structure of {CuO}$_{\textrm{2}}$ sheets and spin-driven high-{Tc}
				superconductivity}},\ }\href {https://doi.org/10.1103/PhysRevB.38.4632}
	{\bibfield  {journal} {\bibinfo  {journal} {Physical Review B}\ }\textbf
		{\bibinfo {volume} {38}},\ \bibinfo {pages} {4632--4659} (\bibinfo {year}
		{1988})}\BibitemShut {NoStop}%
	\bibitem [{\citenamefont {Castro}\ \emph {et~al.}(2006)\citenamefont {Castro},
		\citenamefont {Appel}, \citenamefont {Oliveira}, \citenamefont {Rozzi},
		\citenamefont {Andrade}, \citenamefont {Lorenzen}, \citenamefont {Marques},
		\citenamefont {Gross},\ and\ \citenamefont {Rubio}}]{octopus_local_field}%
	\BibitemOpen
	\bibfield  {author} {\bibinfo {author} {\bibfnamefont {A.}~\bibnamefont
			{Castro}}, \bibinfo {author} {\bibfnamefont {H.}~\bibnamefont {Appel}},
		\bibinfo {author} {\bibfnamefont {M.}~\bibnamefont {Oliveira}}, \bibinfo
		{author} {\bibfnamefont {C.~A.}\ \bibnamefont {Rozzi}}, \bibinfo {author}
		{\bibfnamefont {X.}~\bibnamefont {Andrade}}, \bibinfo {author} {\bibfnamefont
			{F.}~\bibnamefont {Lorenzen}}, \bibinfo {author} {\bibfnamefont {M.~A.~L.}\
			\bibnamefont {Marques}}, \bibinfo {author} {\bibfnamefont {E.~K.~U.}\
			\bibnamefont {Gross}},\ and\ \bibinfo {author} {\bibfnamefont
			{A.}~\bibnamefont {Rubio}},\ }\bibfield  {title} {\emph {\bibinfo {title}
			{octopus: a tool for the application of time‐dependent density functional
				theory}},\ }\href {https://doi.org/10.1002/pssb.200642067} {\bibfield
		{journal} {\bibinfo  {journal} {physica status solidi (b)}\ }\textbf
		{\bibinfo {volume} {243}},\ \bibinfo {pages} {2465--2488} (\bibinfo {year}
		{2006})}\BibitemShut {NoStop}%
	\bibitem [{\citenamefont {Chen}(2006)}]{cuprate_phase_diagram}%
	\BibitemOpen
	\bibfield  {author} {\bibinfo {author} {\bibfnamefont {C.-T.}\ \bibnamefont
			{Chen}},\ }\emph {\bibinfo {title} {Scanning {Tunneling} {Spectroscopy}
			{Studies} of {High}-{Temperature} {Cuprate} {Superconductors}}},\ \href
	{https://doi.org/10.1484/m.art-eb.4.00082} {Ph.D. thesis},\ \bibinfo
	{school} {California Institute of Technology} (\bibinfo {year}
	{2006})\BibitemShut {NoStop}%
\end{thebibliography}
%

\end{bibunit}
\end{document}